%% file: sbnds.tex
\documentclass[aps,prd,preprint,superscriptaddress,nofootinbib]{revtex4-1}

\usepackage{amsmath}
\usepackage{amssymb}
\usepackage{graphicx}
\usepackage{hyperref}
\usepackage[table]{xcolor}
\usepackage{verbatim}
\usepackage[utf8]{inputenc}
\usepackage{multirow}
\usepackage{tikz}
\usetikzlibrary{positioning}
\usetikzlibrary{decorations.pathmorphing,decorations.markings,decorations.shapes,decorations.pathreplacing}

\begin{document}
\tikzset{ 
		scalar/.style={dashed},
		scalar-ch/.style={dashed,postaction={decorate},decoration={markings,mark=at
				position .55 with {\arrow[scale=2]{>}}}},
		fermion/.style={postaction={decorate}, decoration={markings,mark=at
				position .55 with {\arrow[scale=2]{>}}}},
		gauge/.style={decorate, decoration={snake,segment length=0.2cm}},
		gauge-na/.style={decorate, decoration={coil,amplitude=4pt, segment
				length=5pt}}
}

\preprint{PITT-PACC-1906}
\preprint{OSU-HEP-19-07}

\title{Probing the Higgs Portal at the Fermilab Short-Baseline Neutrino Experiments}

\author{Brian Batell}
\email{batell@pitt.edu}
\author{Joshua Berger}
\email{josh.berger@pitt.edu}
\affiliation{PITT PACC, Department of Physics and Astronomy, University of Pittsburgh, Pittsburgh, PA 15260}
\author{Ahmed Ismail}
\email{aismail3@okstate.edu}
\affiliation{PITT PACC, Department of Physics and Astronomy, University of Pittsburgh, Pittsburgh, PA 15260}
\affiliation{Department of Physics, Oklahoma State University, Stillwater, OK 74078}

\date{\today}

\begin{abstract}
The Fermilab Short-Baseline Neutrino (SBN) experiments, MicroBooNE, ICARUS, and SBND, are expected to have significant sensitivity to light weakly coupled hidden sector particles. Here we study the capability of the SBN experiments to probe dark scalars interacting through the Higgs portal. We investigate production of dark scalars using both the Fermilab Booster 8 GeV and NuMI 120 GeV proton beams, 
simulating kaons decaying to dark scalars and taking into account the beamline geometry. We also investigate strategies to mitigate backgrounds from beam-related neutrino scattering events. 
We find that SBND, with its comparatively short ${\cal O}(100\ {\rm m})$ baseline, will have the best sensitivity to scalars produced from the Booster, while ICARUS, with its large detector volume, will provide the best limits on off-axis dark scalar production from NuMI. The SBN experiments can provide leading tests of dark scalars with masses in the 50 - 350 MeV range in the near term. Our results motivate dedicated experimental searches for dark scalars and other long-lived hidden sector states at these experiments. 
\end{abstract}

\maketitle

%%%%%%%%%%%%%%%%%%%%%%%
\section{Introduction      \label{sec:intro}}
\input{intro}

\section{Higgs portal   \label{sec:model}}  
\input{model}

%%%%%%%%%%%%%%%%%%%%%%%
\section{SBN Experiment Setup      \label{sec:SBN}}  
\input{sbn}

%%%%%%%%%%%%%%%%%%%%%%%
\section{Simulation and Analysis      \label{sec:simulation} }
\input{simulation}

%%%%%%%%%%%%%%%%%%%%%%%
\section{Results   \label{sec:results}}  
\input{results}

%%%%%%%%%%%%%%%%%%%%%%%
\section{Outlook \label{sec:outlook}}
\input{outlook}

%%%%%%%%%%%%%%%%%%%%%%%
\appendix*
\section{Simulation validation}
\input{validation}

% If you have acknowledgments, this puts in the proper section head.
\begin{acknowledgments}
We thank Steve Dytman, Vittorio Paolone, Zarko Pavlovic, Gianluca Petrillo and Yun-Tse Tsai for useful discussions and are grateful to Steve Dytman and Yun-Tse Tsai for feedback on the manuscript.
We thank Owen Goodwin and Stefan Soldner-Rembold for pointing out a mistake in the original version of Fig.~\ref{figure:scalar}.
The work of BB and AI is supported in part by the U.S. Department of Energy under grant No. de-sc0007914. We also acknowledge support from PITT PACC. 
We would like to thank the Mainz Institute of Theoretical Physics of the Cluster of Excellence PRISMA+ (Project ID 39083149), the Aspen Center for Physics, which is supported by National Science Foundation grant PHY-1607611, and the Erwin Schr\"odinger International Institute for their hospitality and support during the completion of this project.
\end{acknowledgments}

% Create the reference section using BibTeX:
\bibliography{sbnds}

\end{document}

%% file: intro.tex
% !TEX root = sbnds.tex

Light weakly coupled hidden sectors may play a role in addressing some of the outstanding puzzles in particle physics and cosmology, such as dark matter, neutrino masses, the matter-antimatter asymmetry, the hierarchy problem, and inflation~\cite{Essig:2013lka,Alexander:2016aln,Battaglieri:2017aum,Beacham:2019nyx}.
They also provide an interesting physics target for a variety of intensity frontier experiments. One well known example is proton beam fixed target experiments, including those used to study neutrino oscillations. In these experiments, a beam of relativistic hidden sector particles produced in the primary proton-target collisions passes through a downstream detector and decays to visible Standard Model (SM) particles, providing a distinctive signature that can be discriminated from beam-related neutrino and cosmic-ray induced backgrounds.

The Short-Baseline Neutrino program at Fermilab utilizes three liquid argon time projection chamber detectors -- SBND~\cite{Antonello:2015lea}, 
MicroBooNE~\cite{Acciarri:2016smi}, and ICARUS~\cite{Amerio:2004ze} -- situated along the Booster Neutrino Beam (BNB)~\cite{Machado:2019oxb}.  While the primary physics goals of the SBN program include searches for eV-scale sterile neutrinos (motivated by the LSND~\cite{Athanassopoulos:1996jb} and MiniBooNE~\cite{AguilarArevalo:2010wv,Aguilar-Arevalo:2013pmq,Aguilar-Arevalo:2018gpe} anomalies) and measurements of neutrino-nucleus scattering cross sections, these experiments are also expected to have significant sensitivity to hidden sector particles in the MeV-GeV range. In this paper we study the capability of the SBN experiments to search for dark scalar particles, $S$, interacting via the Higgs portal. 

We consider production of dark scalars from collisions of both the 8 GeV Booster and the 120 GeV Main Injector protons. Concerning the latter, MicroBooNE and ICARUS are located approximately 8$^\circ$ and $6^\circ$ off-axis with respect to the NuMI beam, implying that the associated flux of dark scalars passing through these detectors can be substantial. We perform a careful simulation of the BNB and NuMI beamlines that includes the targets, magnetic focusing horns, and absorbers. 
While we also attempt to take account of various experimental factors such as particle identification and reconstruction efficiencies, detection thresholds, and measurement resolutions, a complete characterization of these parameters is still under active experimental study, so our assumptions in this regard must be considered preliminary. We consider two primary dark scalar signal channels in detail, 1) the $S\rightarrow e^+ e^-$ channel, relevant for scalars below the di-muon threshold, and 2) the combined $S\rightarrow \mu^+\mu^-, \pi^+ \pi^-$ channels which are important for somewhat heavier scalars. We study the beam-related backgrounds to these signatures and design search strategies to efficiently separate them from the signal. We also discuss the potential role of timing measurements as a discriminator for heavy ${\cal O}(100 \,{\rm  MeV})$ scalar particles. 

We find that the SBN experiments have significant sensitivity to dark scalars with masses below a few hundred MeV. 
In particular, SBND, being in close proximity to the BNB target, has the best sensitivity to scalar particles originating from the Booster beam. On the other hand, for scalars produced along the NuMI beamline from the Main Injector, we find that ICARUS, with its large detector volume, will have the leading sensitivity. Notably, kaons decaying at rest (KDAR) after stopping in the NuMI absorber could produce monoenergetic scalars entering ICARUS at a different angle from neutrinos produced in the target, providing a novel signature. In both cases, our projections cover regions of scalar mass -- mixing angle parameter space extending beyond existing experimental limits. Our results motivate dedicated searches for dark scalars and other light weakly coupled particles at the SBN experiments.  
We note that previous studies have examined the sensitivity of the SBN experiments to heavy neutral leptons produced with the Booster beam~\cite{Ballett:2016opr} and the dark-trident signature from dark matter produced in the NuMI beam~\cite{deGouvea:2018cfv}. Furthermore, a number of studies have explored the reach of accelerator-based neutrino experiments to long-lived and other exotic particles such as dark matter in recent years; see e.g. Refs.~\cite{Batell:2009di,Schuster:2009au,Bezrukov:2009yw,Essig:2010gu,Pospelov:2011ha,deNiverville:2011it,Gninenko:2012eq,deNiverville:2012ij,Clarke:2013aya,Blumlein:2013cua,Morrissey:2014yma,Dolan:2014ska,Batell:2014yra,Soper:2014ska,Dobrescu:2014ita,Kahn:2014sra,deNiverville:2015mwa,Gardner:2015wea,Dobrich:2015jyk,Coloma:2015pih,deNiverville:2016rqh,Frugiuele:2017zvx,Darme:2017glc,Berlin:2018tvf,Magill:2018jla,Berlin:2018pwi,deNiverville:2018hrc,Magill:2018tbb,Jordan:2018gcd,deNiverville:2018dbu,Bertuzzo:2018itn,Kelly:2018brz,Kelly:2019wow,Harnik:2019zee,Harland-Lang:2019zur,DeRomeri:2019kic,Ballett:2019bgd,Aguilar-Arevalo:2017mqx,Aguilar-Arevalo:2018wea} and the recent white paper~\cite{Arguelles:2019xgp}.

The outline of this paper is as follows: In Section~\ref{sec:model} we provide a brief review of scalars interacting through the Higgs portal. Section~\ref{sec:SBN} provides an overview of the SBN facillities. 
Section~\ref{sec:simulation} describes our simulation pipeline for the signal and beam related backgrounds, as well as our analysis strategy for the channels under consideration. Our results are presented in Section~\ref{sec:results}, and we conclude in Section~\ref{sec:outlook} with some discussion and outlook. An appendix contains some results of our simulation validation.

%% file: model.tex
% !TEX root = sbnds.tex

We extend the SM by a new singlet scalar particle, $S$, which interacts through the Higgs portal. There are two possible renormalizable portal couplings in general, 
\begin{equation}
-{\cal L} \supset (A S +\lambda S^2) H^\dag H  .
\label{eq:Higgs-portal}
\end{equation}
We will assume the dark scalar acquires a small coupling to SM particles via mass mixing with the Higgs. This always occurs if the super-renormalizable term $A$ in Eq.~(\ref{eq:Higgs-portal}) is nonvanishing. However, even if $A = 0$, the dark scalar can acquire a vacuum expectation value for appropriate choices of the scalar potential, which in turn leads to mass mixing. After diagonalization, there are just two  parameters relevant for our phenomenological study, namely the physical mass $m_S$ of the dark scalar and the scalar-Higgs mixing angle $\theta$:
\begin{equation}
{\cal L} \supset -\frac{1}{2}\,m_S^2 S^2 +  \sin\theta \,   S \, \left( \frac{2m_W^2}{v} W_\mu^+ W^{\mu+} + \frac{m_Z^2}{v} Z_\mu Z^{\mu} - \sum_f \frac{m_f}{v} \bar f f    \right).
\label{eq:L-S}
\end{equation}
Given the tight experimental constraints on the mixing angle, we will be working in the regime $\theta \ll 1$. 
Additional interactions between the scalar and the Higgs (e.g., $h S S$), while important for higher energy collisions at the LHC, will not be relevant for the SBN experiments. As $S$ inherits its couplings in Eq.~(\ref{eq:L-S}) to SM particles from the Higgs, many important phenomenological considerations can be recycled from light Higgs boson studies carried out decades ago~\cite{Gunion:1989we}. 

\begin{figure}[t]
\begin{center}
\includegraphics[width=0.49\textwidth]{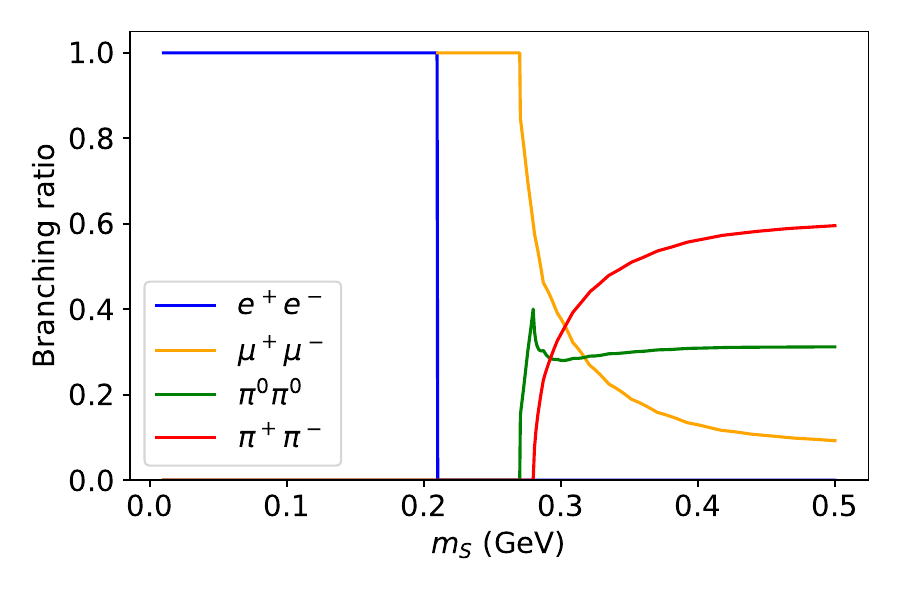}~
\includegraphics[width=0.49\textwidth]{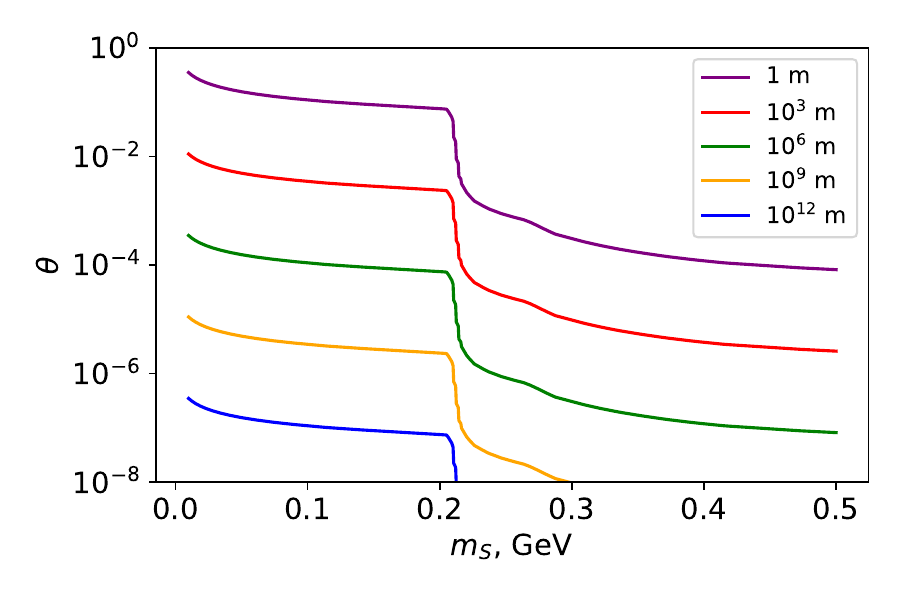}
\caption{
{\it Left}: Scalar branching ratios. {\it Right}: Isocontours of scalar decay length in m in the $m_S - \theta$ plane.
}
\label{figure:scalar}
\end{center}
\end{figure}

Starting from the Lagrangian (\ref{eq:L-S}) we can compute the decay rates of the dark scalar. The partial width to charged leptons is given by
\begin{equation}
\Gamma(S\rightarrow \ell^+ \ell^-) = \theta^2 \, \frac{m_\ell^2 m_S}{8 \pi v^2} \left(1-\frac{4 m_\ell^2}{m_S^2}\right)^{3/2}.
\end{equation}
For $m_S > 2 m_\pi$ hadronic decays of the scalar become important. The description of hadronic decays is complicated by strong interactions and resonance effects leading to sizable theoretical uncertainty in the decay width for scalar masses of order 1 GeV. The total two pion decay for $\pi^0 \pi^0$ and $\pi^+ \pi^+$ can be described as\footnote{Results for scalars with general couplings to SM particles can be found in Ref.~\cite{Batell:2018fqo}.}~\cite{Voloshin:1985tc,Donoghue:1990xh}
\begin{equation}
\Gamma(S\rightarrow \pi\pi) =\theta^2 \frac{3 |G_{\pi}(m_S^2)|^2}{32 \pi \, v^2 \, m_S}  \left(1-\frac{4 m_\pi^2}{m_S^2}\right)^{1/2},
\end{equation}
where we follow the treatment in Ref.~\cite{Feng:2017vli} for the form factor $G_\pi(s)$, which uses the results of Ref.~\cite{Bezrukov:2013fca}. The individual pion channels follow an approximate isospin ratio of $\Gamma(S \to \pi^+ \pi^-) = 2 \Gamma(S \to \pi^0\pi^0)$.  At low energies, the form factor is given by chiral perturbation theory, $G_\pi(s) = \tfrac{2}{9} s+  \tfrac{11}{9}  m_\pi^2$, but this model differs from the full numerical calculation we use. Within the model described by Eq.~\eqref{eq:L-S}, any other SM decays have negligible branching fraction. In particular, there are no other dark sector states into which the scalar $S$ might decay. In Figure~\ref{figure:scalar} we show the scalar branching ratios as a function of its mass, as well as isocontours of the scalar decay length in the $m_S - \theta$ plane.

%
%%%%%%%%%% FIGURE %%%%%%%%
%%%%%%%%%%%%%%%%%%%%%%%%%%

\begin{figure}[!htb]
\centering
\begin{tikzpicture}
\coordinate (mp) at (6.4,-4);
\coordinate[left=0.75cm of mp] (pvi);
\coordinate[left=1.5cm of pvi] (s);
\coordinate[below=1.05cm of pvi] (spec);
\coordinate[left=1.5cm of spec] (qi);
\coordinate[right=3cm of spec] (qf);
\coordinate[right=1.5cm of pvi] (pvf);
\coordinate[right=1.5cm of pvf] (d);
\coordinate (Sf) at ([shift={(30:1.5cm)}]mp);
\coordinate[left=0.5cm of s] (lbt); 
\coordinate[left=0.5cm of qi] (lbb);
\coordinate[right=0.5cm of d] (rbt); 
\coordinate[right=0.5cm of qf] (rbb);
\coordinate (dp) at ([shift={(45:1.5cm)}]Sf);
\coordinate (dm) at ([shift={(15:1.5cm)}]Sf);

\draw[fermion] (s) node[left=0.1cm,font=\fontsize{12pt}{12pt}\selectfont] {$s$} -- (pvi);
\draw[fermion] (pvi)  -- (mp) node[below=0.1cm,font=\fontsize{12pt}{12pt}\selectfont] {$c,t$};
\draw[fermion] (mp)  -- (pvf);
\draw[gauge] (pvf) arc (0:-180:0.75cm) node[pos=0.75,left=0.3cm,font=\fontsize{12pt}{12pt}\selectfont] {$W$};
\draw[scalar] (mp) -- node[above left=0.1cm,font=\fontsize{12pt}{12pt}\selectfont] {$S$} (Sf);
\draw (Sf) -- (dp) node [right=0.1cm] {$\ell,\pi$};
\draw (Sf) -- (dm) node [right=0.1cm] {$\ell,\pi$};
\draw[fermion] (qf) node[right=0.1cm,font=\fontsize{12pt}{12pt}\selectfont] {$q$} -- (qi) node[left=0.1cm,font=\fontsize{12pt}{12pt}\selectfont] {$q$};
\draw[fermion] (pvf) -- (d) node[right=0.1cm,font=\fontsize{12pt}{12pt}\selectfont] {$d$};
\draw[decorate,decoration={brace,amplitude=5pt}] (lbb) -- node[left=0.4cm,font=\fontsize{12pt}{12pt}\selectfont] {$K$} (lbt);
\draw[decorate,decoration={brace,amplitude=5pt}] (rbt) -- node[right=0.4cm,font=\fontsize{12pt}{12pt}\selectfont] {$\pi$} (rbb);
\end{tikzpicture}
\caption{Dominant signal production process via kaon decays.}\label{fig:signal-diagram}	
\end{figure}
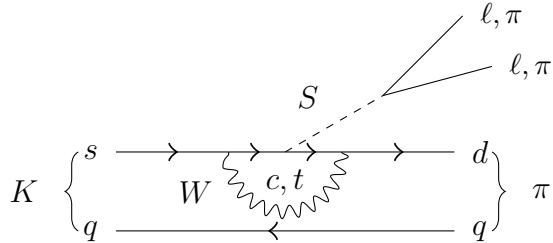

For the SBN experiments the most important production channel will be through kaon decays, $K\rightarrow \pi S$, which proceed via the one-loop penguin and scalar bremsstrahlung processes~\cite{Willey:1982mc,Leutwyler:1989xj,Gunion:1989we,Bezrukov:2009yw,Feng:2017vli}. The dominant contribution comes from the $W$ boson - top quark penguin shown in Fig.~\ref{fig:signal-diagram}, leading to the partial decay width,  
\begin{equation}
\Gamma(K^\pm \rightarrow \pi^\pm S) \simeq\frac{ \theta^2}{16 \pi m_K} \bigg\vert \frac{3 V^*_{td} V_{ts} m_t^2 m_K^2 }{32 \pi^2 v^3}  \bigg\vert^2  \lambda^{1/2}\left(1, \frac{m_S^2}{m_K^2}, \frac{m_\pi^2}{m_K^2}\right),
\end{equation} 
with $\Gamma(K^0_L \rightarrow \pi^\pm S) \simeq \Gamma(K^\pm \rightarrow \pi^\pm S)$. The branching ratio for scalars produced in $K^0_S$ decays is smaller by several orders of magnitude compared to those from $K^\pm$ and $K^0_L$, owing to suppression from the small CP-violating phase and the substantially shorter $K^0_S$ lifetime.

Besides kaon decays, $B$ mesons have an even larger branching ratio to $S$ and furthermore allow for production of heavier scalars~\cite{Grinstein:1988yu,Chivukula:1988gp,Batell:2009jf,Bezrukov:2009yw,Clarke:2013aya}. However, while such decays are important for higher energy facilities such as the SPS and LHC beams at CERN, they will not be relevant for the lower energy Booster and Main Injector beams due to the substantially lower $B$ production rate. A more promising process at these energies is scalar production through proton bremsstrahlung, which also allows for production of heavier scalars.  We have estimated the sensitivity from proton bremsstrahlung using the results of Ref.~\cite{Boiarska:2019jym}, finding that it leads to no additional sensitivity beyond that from Kaon decays. 
Additional sources of low mass scalar production at these energies include rare $\pi^\pm,\eta,\eta'$ decays, although they are subdominant to kaon decays and will be neglected in our study.

There are a number of experimental and astrophysical constraints on the Higgs portal scalar parameter space, but we will defer a discussion of these to Section~\ref{sec:results} below. We now move on to discuss the SBN experiments. 

%% file: sbn.tex
% !TEX root = sbnds.tex

\begin{figure}[h]
\centering
\begin{tabular}{c c c}
\includegraphics[height=2.75in]{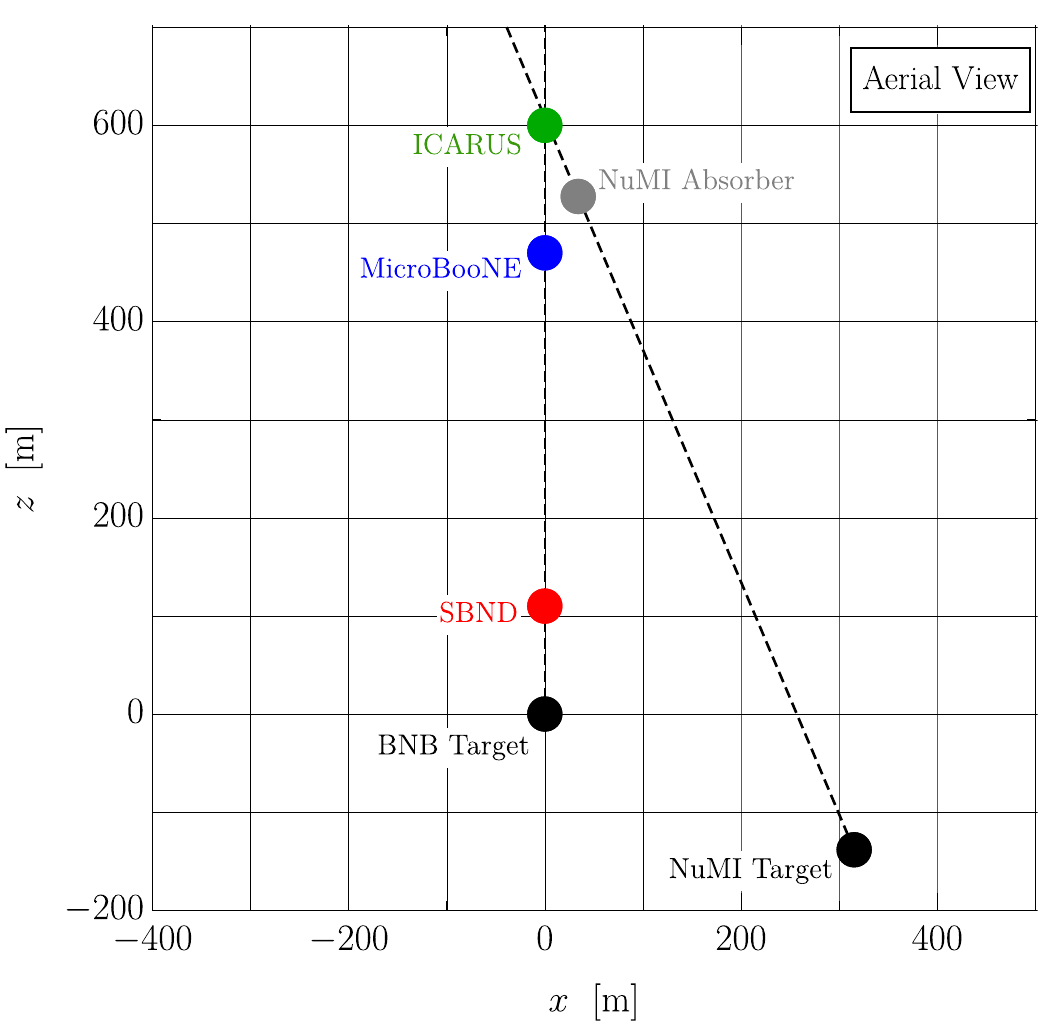} &  ~ &
\includegraphics[height=1.75in]{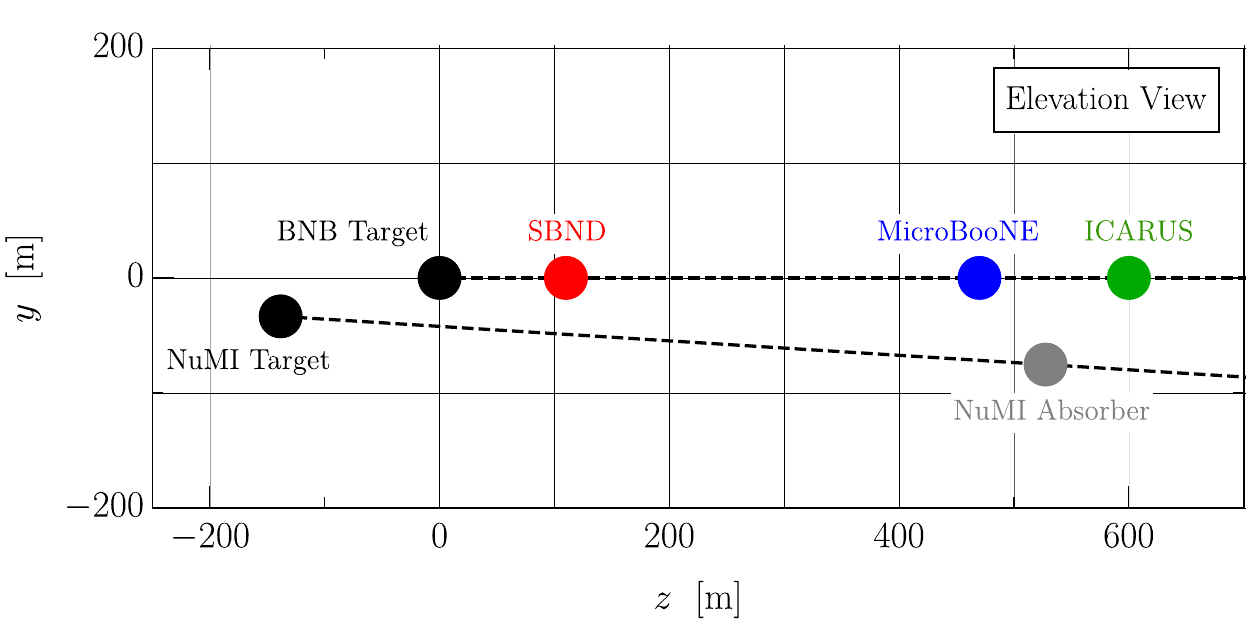}
\end{tabular}
\caption{ Coordinates in the ``BNB frame'' defined by a left-handed coordinate system with the origin at the BNB target, the $z$-axis pointing to center of the SBN detectors, and the $y$-axis pointing vertically upwards. We show the coordinates of the BNB and NuMI targets, NuMI absorber, and the SBND, MicroBooNE, and ICARUS detectors in both aerial (left) and elevation (right) views.} 
\label{fig:map1}
\end{figure}

The sensitivity of an experiment to decays of long-lived dark sector particle decays depends primarily on the number of such particles that can be made to pass through the detector and the ability of the detector to distinguish the decays from backgrounds.  In both these regards, the Short-Baseline Neutrino (SBN) program at Fermilab offers strong capabilities for near future sensitivity.  For Higgs portal scalars the dominant production mechanism comes from kaon decays.
The Booster Neutrino Beam (BNB) and the Neutrinos at the Main Injector (NuMI) beam offer a large production of both focused and stopped kaons with several detectors in the 100 to 1000 m range of their targets.  Among these detectors, the Liquid Argon Time Projection Chamber (LArTPC) detectors of the SBN program, namely the Short-Baseline Near Detector (SBND), MicroBooNE, and ICARUS-T600, will have low thresholds and excellent particle identification. We therefore focus our study on the signals of beam-produced dark scalars decaying in these detectors. Figure~\ref{fig:map1} shows aerial and elevation views of the Fermilab campus, including the locations of the BNB and NuMI targets, NuMI absorber, and the SBND, MicroBooNE, and ICARUS detectors.
In this section, we discuss the capabilities of both the beams and SBN detectors in turn.

\subsection{Beams}

\begin{table}[!htb]
	\centering
	\begin{tabular}{l c c}
		\hline\hline
		Beam & BNB & NuMI \\
		\hline
		Kinetic Energy, GeV & 8 & 120 \\
		Target & Beryllium & Graphite \\
		Dist.~to SBND, m & 110 & 401\\
		Angle w.r.t.~SBND, rad & 0 & 0.53 \\
		Dist.~to MicroBooNE, m & 470 & 685 \\
		Angle w.r.t.~MicroBooNE, rad & 0 & 0.14 \\
		Dist.~to ICARUS-T600, m & 600 & 803 \\
		Angle w.r.t.~ICARUS-T600, rad & 0 & 0.097 \\ 
		\hline\hline
	\end{tabular}
%\end{comment}
	\caption{Summary of the properties of the current Fermilab neutrino beams. } 
	\label{tab:beams}
\end{table}
The Fermilab accelerator neutrino program currently consists of the aforementioned BNB and NuMI beams.  Each has its own advantages and disadvantages for the purposes of dark scalar searches and will have its best sensitivity at different SBN detectors.  We therefore discuss the two in turn. 
Their properties are summarized in Table \ref{tab:beams}.

\subsubsection{BNB}

The BNB consists of 8 GeV kinetic energy protons impinging on a beryllium target while traveling close to due North. The SBN detectors are lined up along the beam axis at distances from 110 m (SBND) to 600 m (ICARUS-T600). The accelerator delivered $3.1 \times 10^{20}$ protons on target (POT) in the 2018 fiscal year~\cite{FNAL:2018}, for a current total of around $1.3 \times 10^{21}$ POT during MicroBooNE operation and an equivalent roughly $1.3 \times 10^{21}$ expected from running for around 4 years at that intensity for SBND and ICARUS.  We take this number, $1.3 \times 10^{21}$, as our benchmark for our BNB-based studies.  
Charged mesons are focused by a magnetic horn to deliver a beam enhanced in either neutrinos or anti-neutrinos.

\subsubsection{NuMI Beam}

The NuMI beam is made by striking 120 GeV kinetic energy protons on a graphite target.  The beamline is roughly north pointing along the axis from Fermilab to the MINOS far detector in Minnesota.  The SBN detectors are not along the beam axis, but are slightly off axis.  The beam does however pass at small angle with respect to both MicroBooNE and ICARUS-T600.  Given the higher energy of the beam and the angular spread in the produced scalars, these detectors can have significant sensitivity to scalars produced with the NuMI beam.  The MicroBooNE detector is located a distance of roughly 685 m from the target at an angle of $0.14$ radians ($7.8^\circ$), while the ICARUS-T600 is 803 m at an angle of $0.097$ radians ($5.5^\circ$).  The accelerator delivered $5.7\times 10^{20}$ POT in the 2018 fiscal year~\cite{FNAL:2018}, for a projected total of close to $3 \times 10^{21}$ expected from running for 5 years, which we take as our benchmark for NuMI-based studies.   
The NuMI beam has two horns whose position and current can be adjusted depending on the desired energy and type (neutrino or antineutrino) of the beam. The actual running mode will be determined by NO$\nu$A physics requirements. However, because an order-1 fraction of our signal will come from decays of neutral mesons which are unaffected by the magnetic horns, we do not expect the mode to significantly affect the final results.  We assume the standard ``medium energy'' mode horn locations, which have been used for the past several years~\cite{Adamson:2015dkw}, and neutrino mode running with a 200 kA forward horn current configuration below.  

\subsection{SBN Detectors}

The SBN program consists of the SBND, MicroBooNE and ICARUS-T600 LArTPC detectors positioned along the BNB.  For the most part, the detectors function in similar manners, with similar TPC designs.  Though we assume the same detection capabilities for the three SBN detectors in this study, we do account for their difference in volume and location, as outlined in Table \ref{tab:detectors}.  MicroBooNE is currently operating, while SBND and ICARUS-T600 are expected to be collecting data by the end of 2019 and beginning of 2021 respectively~\cite{FNAL-SBN,yuntse-private}. The volume and mass are particularly relevant for signal and background rate calculation respectively as described below.  Note that while we list the full dimensions for a box shaped detector volume, we make the simplifying approximation that the rates for the signal do not depend on the detailed geometry of the detector and perform our calculations using a spherical detector with the correct fiducial volume.  This approximation is valid in the limit that the detector dimensions are much less than the distance that the scalar travels before decaying.  Since the detector dimensions are $\mathcal{O}(1~{\rm m})$, while the distance the scalar travels is $\mathcal{O}(100~{\rm m})$, this approximation is valid at the $1\%$ level.

We discuss detector effects and event reconstruction in detail below in Section \ref{sec:detector-effects}, after further discussing the relevant signals and backgrounds. 

\begin{table}[!htb]
	\centering
	\begin{tabular}{@{\hspace{0.25cm}}l@{\hspace{0.5cm}}|@{\hspace{0.5cm}}c@{\hspace{0.5cm}}c@{\hspace{0.5cm}}c@{\hspace{0.25cm}}}
		\hline\hline
		Detector & SBND & MicroBooNE & ICARUS-T600 \\
		\hline
		Dimensions ($\text{m}\times\text{m}\times\text{m}$) & $3.67 \times 3.70 \times 4.05$ & $2.26 \times 2.03 \times 9.42$ & $2\times (2.67 \times 2.86 \times 17.00 )$\\
		Mass (tons) &77.0  & 60.5 & 363 \\
		Operation dates & Early 2021 -- 2024 & Operating until 2024 & Late 2019 -- 2024 \\
		\hline\hline
	\end{tabular}
	\caption{Summary of SBN detector properties~\cite{Antonello:2015lea}.  The dimensions and masses assumed are those of the fiducial volume for the $\nu_\mu$ analyses.}\label{tab:detectors}
\end{table}

%% file: simulation.tex
% !TEX root = sbnds.tex

\subsection{Signal event generation} 

To simulate the proton collisions from the Booster and NuMI beams, we employ the \verb+g4bnb+~\cite{AguilarArevalo:2008yp} and \verb+g4numi+~\cite{Aliaga:2016oaz} codes. Each beamline is assumed to be configured in neutrino mode. The simulations, based on \verb+Geant4+~\cite{Brun:1994aa}, incorporate the geometry of each beamline including the targets, focusing horns and decay volumes. They take into account the full evolution of the particles produced in the primary collisions of protons with the target, including not only secondary meson production but also the subsequent interactions with the beamline elements. In both codes, whenever a neutrino is produced in a particle decay, information about the neutrino momentum and position, as well as its ancestor, is stored using the \verb+dk2nu+~\cite{dk2nu} format.

\begin{figure}[t]
\begin{center}
\includegraphics[width=0.49\textwidth]{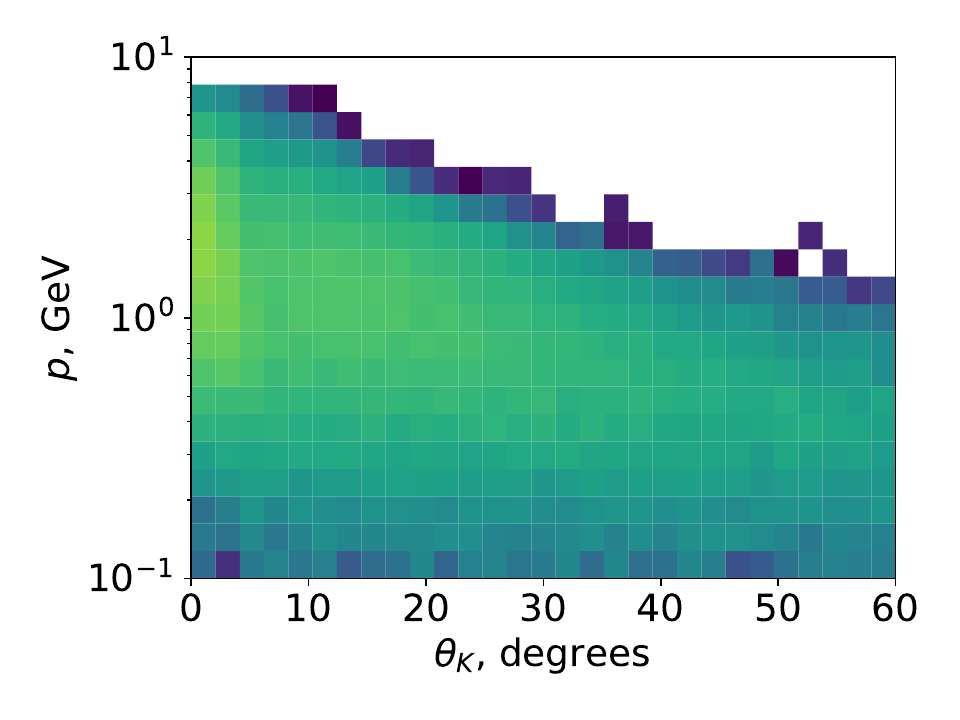}~
\includegraphics[width=0.49\textwidth]{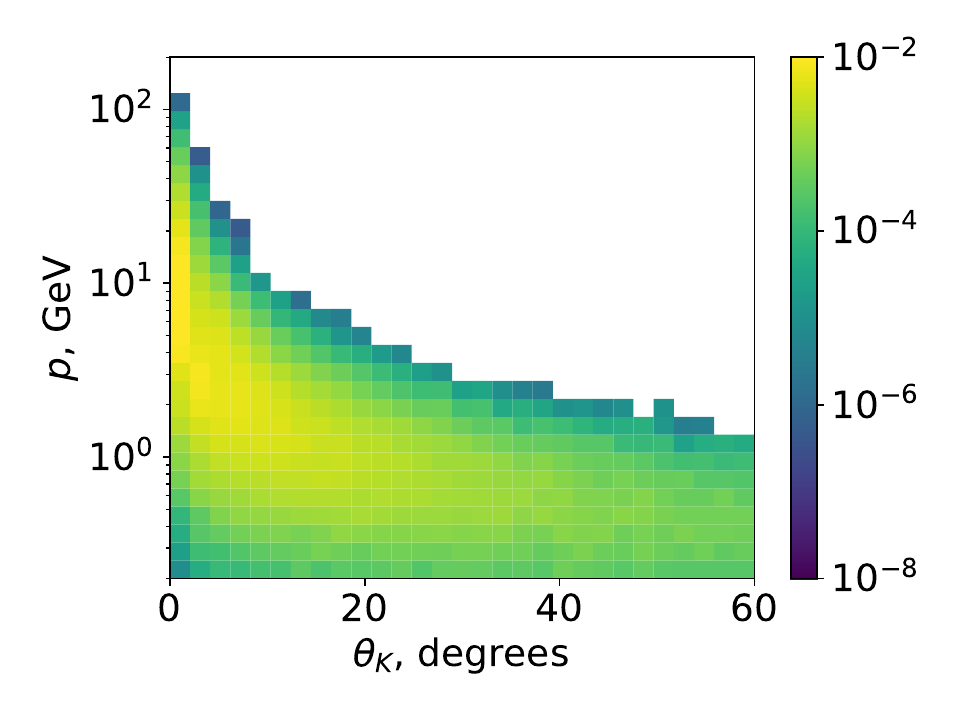}
\caption{
{Kinematics of kaons produced per POT by \texttt{g4bnb} (left) and \texttt{g4numi} (right) simulations, at the time of their decays to neutrinos. The angle is taken with respect to the beamline. Kaons that stop before decaying are not shown above; see the text for details.}
}
\label{fig:kaonkin}
\end{center}
\end{figure}

We use the ancestry information provided in the \verb+g4bnb+/\verb+g4numi+ output to extract the momentum and position of each kaon that decayed to a neutrino, at the time of decay. In Figure~\ref{fig:kaonkin}, we show the kinematics of the kaons from both beams.
Most kaons are produced in the targets, and travel in the direction of the beamlines. The initial kaon momentum is characteristic of the beam energy. Because of subsequent interactions of the kaons with the beam apparatus as well as secondary production, the kinematics of the kaons at the time of decay to neutrinos can differ from the kinematics of the kaons produced in the beam targets. The most notable effect is that at the BNB (NuMI beam), approximately 2/3 (1/3) of the kaons produced are stopped before decaying. These kaons, which decay isotropically in the lab frame, are not shown in Figure~\ref{fig:kaonkin}. For NuMI, the kaons that are stopped in the absorber provide an interesting source of signal, as the resulting scalars are monoenergetic and enter the Booster beamline detectors at an angle that is quite different from those of their counterparts arising from kaons decaying in the target. We will discuss this possibility further In Sec.~\ref{sec:analysis}.

Using the kaons, we then simulate $K \to \pi S$ decays for different scalar masses, yielding the phase space distribution of scalars produced in each beamline. We consider only decays of $K^\pm$ and $K_L^0$, as the $K_S^0$ provides a negligible contribution due to its significantly smaller branching ratio to scalars. The mixing angle $\theta$ sets the overall normalization of the distribution, i.e.~the number of scalars produced per POT.

\begin{figure}[t]
\begin{center}
\includegraphics[height=2.5in]{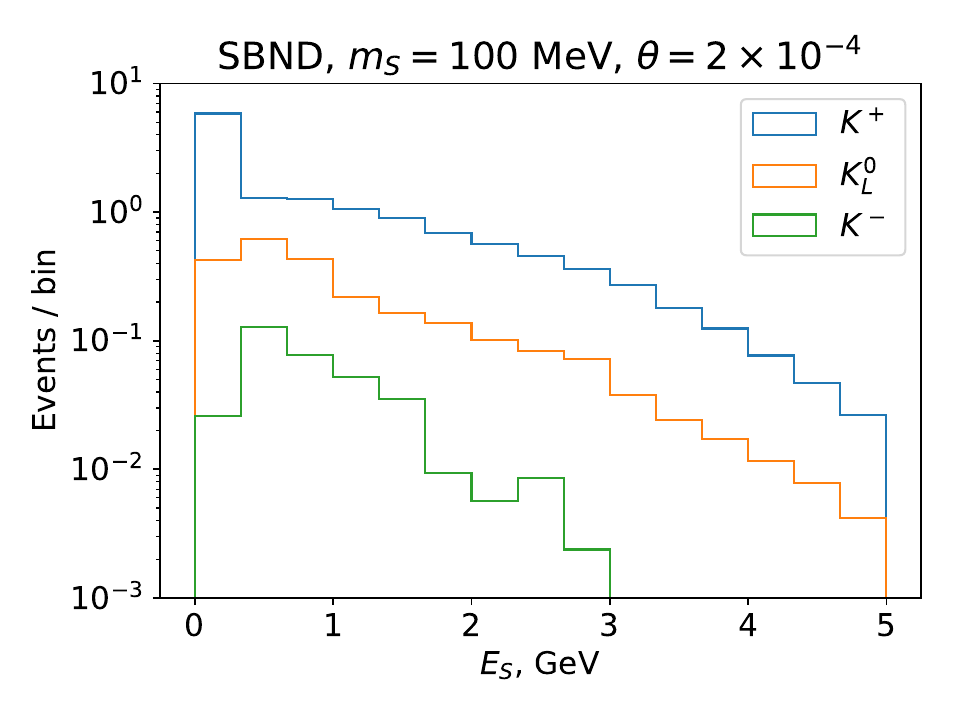}~
\includegraphics[height=2.5in]{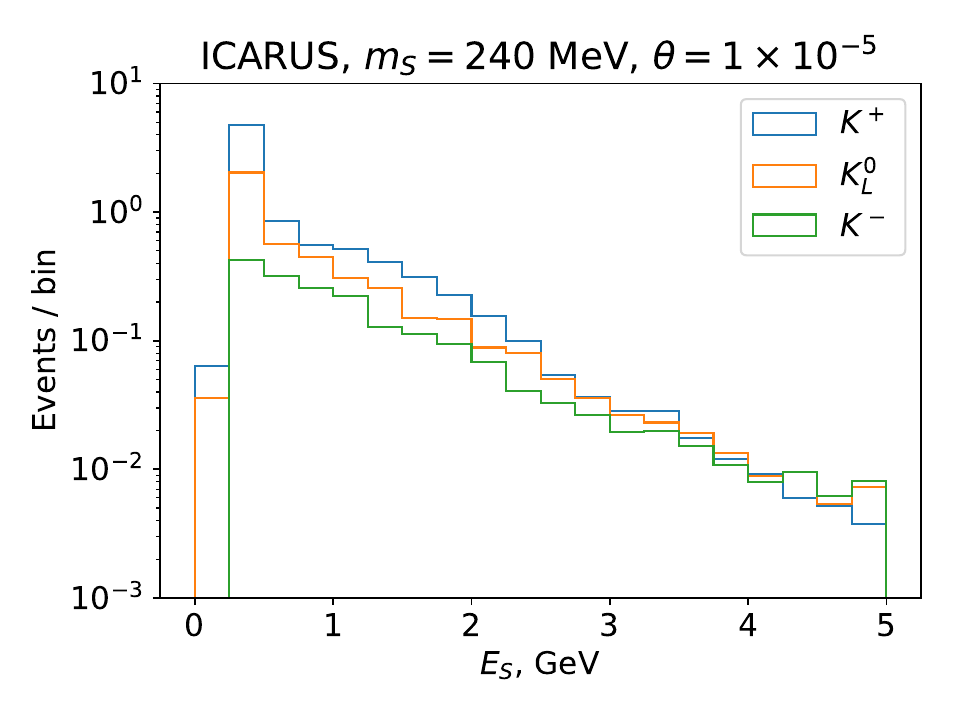}
\caption{
Energies of scalars produced by \texttt{g4bnb} that decay in SBND (left) and by \texttt{g4numi} that decay in ICARUS (right).
There are peaks at low $p$ in the $K^+$ distributions from KDAR.
}
\label{fig:skin}
\end{center}
\end{figure}

Finally, for each of the SBN detectors, we consider the scalars from each beam that would pass through the detector volume. If a scalar with velocity $\beta$ and rest lifetime $\tau$ would enter and exit a given detector at distances $L_{\mathrm{in}}$ and $L_{\mathrm{out}}$ from its production point, the probability that it decays inside the detector is
\begin{equation}
\label{eq:decayprob}
P = e^{-L_{\mathrm{in}} / \gamma \beta \tau} - e^{-L_{\mathrm{out}} / \gamma \beta \tau} .
\end{equation}
Figure~\ref{fig:skin} shows the distributions of the resulting scalar energy at the detectors which see the most decaying scalars from each beam. For on-axis production at the BNB, SBND covers the greatest solid angle by far, and generally has the most scalar decays of the SBN detectors. For off-axis production, both MicroBooNE and ICARUS are approximately 700 m away from the NuMI target and only a few degrees off axis, while SBND is somewhat closer but much further off axis, approximately 30 degrees. Because of the much larger ICARUS volume (see Table~\ref{tab:detectors}), more scalars from NuMI decay in ICARUS than in the other SBN detectors. Typically, the scalar parent is a highly boosted kaon, so the scalar and its decay products are all well-aligned with the beam direction. In Sec.~\ref{sec:analysis}, we use these characteristic features of the scalar decay products to design cuts to reduce backgrounds from neutrinos and cosmic rays.

We have checked our pipeline in two different ways. First, we have reproduced several neutrino flux predictions, at both the SBN detectors as well as MINERvA, with \verb+g4bnb+ and \verb+g4numi+. The results of this validation are in the Appendix. In addition, we have replaced the \verb+Geant4+-based codes with \verb+Pythia8+~\cite{Sjostrand:2014zea} simulations of the primary proton-target collisions, neglecting additional interactions of the secondary particles within the beamline as well as the propagation of the kaons before decaying to scalars. This check yields similar results to those obtained in Sec.~\ref{sec:results} with the dedicated codes.

\subsection{Background Generation}

There are several sources of potential background events for the signal we are considering, though all are in principle reducible.  As the SBN detectors are operated on the surface, cosmic ray induced backgrounds are a significant worry.  On the other hand, it is rare for cosmic rays to mimic the signal we are considering, as any tracks or showers should come in pairs and generally appear in the middle of the detector.  Furthermore, systems are being designed specifically to tag and eliminate cosmic ray backgrounds at MicroBooNE~\cite{Adams:2019bzt}, while SBND and ICARUS are also being designed with dedicated cosmic ray taggers as well~\cite{Antonello:2015lea}.  Beyond such a tagger, a combination of timing and angular cuts can further reduce this background, though it could in principle still hinder reconstruction of signal events.  We do not consider this background further in this study.

The remaining backgrounds are due to beam neutrinos.  Both interactions in the dirt and in the  detector volume can contribute, though we only consider interactions in the fiducial detector volume, which should dominate for the analysis we have developed.  Even this dominant background cannot fully mimic our signal and several cuts can be used to drastically reduce it, as discussed below.

To generate the background, we use the neutrino flux predictions presented in \cite{Adams:2013uaa}.  The fluxes are used for input for generation of neutrino interactions on ${}^{40}\text{Ar}$ using \verb+GENIE v3+.  The $\pi^0$ in the \verb+GENIE+ output are decayed to photons.  These photons, along with any other final state particles reported to \verb+GENIE+ are smeared and thresholds are applied according to the description in Section \ref{sec:detector-effects} below.  

\subsection{Detector Effects \label{sec:detector-effects}}

In this section, we discuss our assumed detection thresholds, energy and angular resolutions, and particle identification capabilities.

The target volume consists of liquid argon, which provides a dense medium for neutrino interactions.  This density provides an increased worry about neutrino-generated backgrounds, but the other features make the SBN detectors nevertheless well-suited to a search for decaying dark sector particles.  The liquid argon is placed in an electric field which drifts ionized electrons due to energetic charged particles traveling through the argon.  The electrons are collected on wire planes that allow for roughly millimeter precision measurement of the position of the interaction in the plane direction.  The third coordinate of the charge deposition is reconstructed based on the drift time.  The time of the interaction can also be reconstructed with nanosecond precision if scintillation light from the interaction is collected.  The deposited energy can be determined from the charge collection, allowing a reconstruction of the energy deposited by a track per unit length, which in turn allows for strong particle identification and calorimetry.  Given all of this information, the full four vector of the particles produced through an interaction in the detector volume can be reconstructed with excellent angular resolution and good energy resolution.

Many of the capabilities of the SBN LArTPC detectors are under active investigation and are not known.  For the purposes of our study, we assume certain figures of merit that we believe are reasonable.  As understanding of the detectors improves, we expect to have updated sensitivities.  We now discuss the properties that we assume, as well as some motivation for these assumptions.

Particle identification is one of the main advantages of LArTPC technology.  That said, certain particles can be rather tricky to distinguish.  The particles that are stable on the scale of the position resolution of LArTPC detectors in the 100 MeV energy range of interest are $e^\pm$, $\gamma$, $p$, $\mu^\pm$,  and $\pi^\pm$.  The first two are generally referred to as shower particles as they create electromagnetic cascade showers above $33~\text{MeV}$, while the last three are referred to as track particles, as they appear as single tracks.  It is worth noting that, in some instances, the distinction between track and showering particles is not 100\% accurate.  We assume however that this distinction is perfectly accurate in the analysis below.

The signal of interest consists of an isolated $e^+e^-$, $\mu^+\mu^-$ or $\pi\pi$ pair.  For the $\pi\pi$ case, we focus on the simpler and more prevalent $\pi^+\pi^-$ topology, though a $2\pi^0$ decay is also possible. Since no neutrino interactions from the beam produce precisely this final state, the backgrounds are, in principle, reducible.  There are, however, neutrino interactions that can fake our signal in a LArTPC.  

For $e^+e^-$ pairs, the primary worry is from photons arising from either neutral pion decay or hard radiation.  Photons and electrons both lead to electromagnetic showers in the detector.  Photons, however, only begin their shower after a conversion length of roughly $14~{\rm cm}$.  Both single and two photon events can fake the signal absent any other visible activity from the neutrino interaction.  Single photons will convert into an $e^+e^-$ pair, which directly mimics our signal.  The kinematics of the conversion process, however, very rarely lead to widely separated $e^+e^-$ pairs and the pair typically has small invariant mass.  Based on \verb+Geant4+ simulations, we find that $92.5\%$ of $100~{\rm MeV}$ photon conversions and $99\%$ of $500~{\rm MeV}$ photons conversions lead to electron-positron pairs with an opening angle of less than $10^\circ$.  This background is thereby virtually eliminated.   The isolation cut also effectively eliminates the two photon background.  The signal showers should begin with two electron tracks that originate at the same spatial point.  The two photon background, however, is expected to have its showers starting at well-separated vertices due to the differing conversion lengths of the two photons.  In fact, for isolated photons, a straightforward calculation based on the conversion length of 14 cm shows that over $99.5\%$ of photon pairs have vertices separated by more than the position resolution of LArTPCs, which we take to be $3~{\rm mm}$.  Similar arguments eliminate the subdominant backgrounds due to charged-current electron production in association with a photon.  Therefore, for a sufficient isolation cut, the search for $e^+e^-$ pairs should be effectively background free.

For the remaining searches, the primary background comes from pions and muons.  Protons, muons, and charged pions all leave tracks, but protons are distinguishable by their distinct energy deposit per unit length compared with track length.  It is worth noting that some of the muons and pions from our signal have large energies and are expected to not be contained by the SBN detectors.  From \verb+Geant4+ simulations, we expect a muon of energy $1~{\rm GeV}$ to have a track length of around $4~{\rm m}$, for example.  A distribution of muon candidate energies will be shown below in Fig.~\ref{fig:maxmu-mumu}.  This does not necessarily mean that the event cannot be reconstructed, but rather makes reconstruction more challenging.  For the present analysis, we do not take containment effects into account though we do not expect this to significantly affect our sensitivity estimates.  Muons and charged pions are in general quite challenging to tell apart as they deposit energy at similar rates.  Charged pions leave larger energy deposits at the end of their tracks due to their additional nuclear interactions, but the ability to utilize this has not yet been quantified.  We therefore conservatively assume that they are indistinguishable.  Note that the charge of the muon and pion should be distinguishable based on capture of $\mu^-$ that is not possible for the opposite charge. The effectiveness of this procedure has not yet been determined, so we do not attempt to use the muon or pion charges. Since pions and muons are assumed to be indistinguishable, the backgrounds to searches are typically somewhat large, coming from $\mu^\pm\pi^\mp$ inelastic charge current events and $\pi^\pm\pi^\pm$ deep inelastic events. We make use of kinematic cuts as described below to reduce these backgrounds.

Next, we discuss the kinetic energy thresholds for observing particles. Thresholds in LArTPCs are typically in the 10-100 MeV range, though they depend on the specific detector, context and particle.  For on-beam analyses, the threshold is set by requiring that a sufficient number of hits are seen in the collection plane.  MicroBooNE typically requires at least 20, setting thresholds of roughly 80 MeV for protons, 40 MeV for $\pi^\pm/\mu^\pm$, and 20 MeV for EM showers, though muons are generally required to have longer tracks.  Reconstruction of shorter proton tracks should be possible as demonstrated by ArgoNeuT~\cite{Acciarri:2018myr}, allowing for thresholds of 21 MeV for protons and 10 MeV for $\pi^\pm/\mu^\pm$.  The DUNE CDR quotes a comparable set of thresholds used for its Fast Simulation~\cite{Acciarri:2015uup}. 
Light collection is crucial for off-beam analyses and helps with energy reconstruction in all cases.  We assume that light collection thresholds are met if the tracking thresholds are met.

We assume a threshold of $30~{\rm MeV}$ for both electrons and photons, motivated by the requirement that they have at least one hard interaction in the detector.  We also assume a threshold of $30~{\rm MeV}$ for muons and pions, as motivated by a requirement of at least 10 hits in the TPC. While this is slightly more aggressive than the current MicroBooNE assumption, improvements on reconstruction should be made in the near future.  We also reconstruct protons with a threshold of $20~{\rm MeV}$ for the purposes of vetoing.  While this is lower than the nominal reconstruction threshold, it is noted that proton reconstruction is not required; we only need to determine whether a proton was present in the event.

Energy resolution is key for making an accurate mass determination of the decaying scalar. For its Fast Monte Carlo, the DUNE CDR quotes 30\% for exiting muon-like tracks, with better resolutions for contained tracks based on track length~\cite{Acciarri:2018myr}.  They further quote a resolution of $2\% \oplus 15\%/\sqrt{E}~[{\rm GeV}]$ for electromagnetic showers.  A phenomenological study of heavy neutral leptons adopted yet different numbers, $6\%/\sqrt{E}~[{\rm GeV}]$ for muons and $15\%/\sqrt{E}~[{\rm GeV}]$ for electromagnetic showers~\cite{Ballett:2016opr}.  A recent MicroBooNE thesis has found even better resolution for contained muons and good resolution on the neutral pion peak~\cite{MicroBooNE:2018,DelTutto:2019cpu}.  A study of momentum reconstruction for muons from multiple Coulomb scattering, suitable for exiting muons found resolution in the 10-25\% range across different energies~\cite{Antonello:2016niy}. It is worth noting that some of the region of interest for our searches involves high energy particles that may not be contained in the detector, which would degrade both resolution and particle identification capabilities.  Based on these numbers, we assume $10\%$ resolution on electromagnetic showers, $3\%$ momentum resolution on muons and pions below $300~{\rm MeV}$ kinetic energy as these are likely to be contained, and $10\%$ momentum resolution on muons and pions above this threshold as they are likely to exit. A lower resolution would require a wider mass window around the candidate scalar mass, increasing the background somewhat, but not drastically changing our results.

Finally angular resolution should be excellent for tracks, given the excellent position resolution.  We adopt a universal angular resolution for particles of 0.03 rad or $1.73^\circ$ as in Ref.~\cite{Antonello:2015lea}.

\begin{table}[!htb]
	\centering
	\begin{tabular}{@{\hspace{0.5cm}}l@{\hspace{0.5cm}}|@{\hspace{0.5cm}}c@{\hspace{0.5cm}}c@{\hspace{0.5cm}}}
		\hline\hline
		Particle & Threshold (MeV) & Energy Resolution \\
		\hline
		$e^\pm$/$\gamma$ & 30 & 10\%\\
		\multirow{2}{*}{$\mu^\pm$/$\pi^\pm$} & \multirow{2}{*}{30} & 3\% (${\rm KE} < 300~{\rm MeV}$) \\
		 &  & 10\% (${\rm KE} > 300~{\rm MeV}$) \\
		$p$ & 20 & -- \\
		\hline\hline
	\end{tabular}
	\caption{Summary of the detector effects assumed in this study.  The thresholds for the electromagnetic and muon-like particles are in line with the best available studies thus far~\cite{Acciarri:2015uup}.  
	The resolutions are adapted from Refs.~\cite{MicroBooNE:2018,DelTutto:2019cpu}.}\label{tab:reco}
\end{table}
Our analysis assumptions are shown in Table \ref{tab:reco}.  We apply these thresholds and smearing at the level of final state four vectors from the event generator chains, noting that these numbers are subject to change as LArTPCs are better understood.  We further veto on additional photons and electrons above 2 MeV, as well as any additional pions or muons, which will also lead to visible electrons when they decay.

We make one final comment regarding neutrons and recoiling nuclei.  Both can lead to additional energy deposits in the detector, the neutrons scattering widely throughout the volume~\cite{Friedland:2018vry} and the recoiling nucleus activity centered near a neutrino interaction point~\cite{yuntse-private}. Decaying mediators do not leave such signals in the detector, so vetoing such activity would be an excellent way to reduce backgrounds.  Nevertheless, there is no official study of these effects, so we simply ignore neutrons and recoiling nuclei in our analyses.

\subsection{Analysis   \label{sec:analysis}}

Below, we describe our analysis strategy for the various signal channels. 

\subsubsection{$e^+ e^-$}
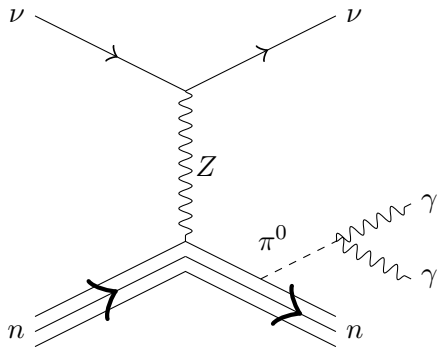
\begin{figure}[!htb]
	\begin{tikzpicture}
	\draw[fermion] (0,0) node[left] {$\nu$} -- (2,-1);
	\draw[fermion] (2,-1) -- (4,0) node[right] {$\nu$};
	\draw[gauge] (2,-1) -- node[right] {$Z$} (2,-3);
	\draw (0,-4) -- (2,-3) -- (4,-4);
	\draw[postaction={decorate}, decoration={markings,mark=at
		position .55 with {\arrow[scale=5]{>}}}] (0,-4.2) node[left] {$n$} -- (2,-3.2);
	\draw (2,-3.2) -- (3,-3.7);
	\draw[postaction={decorate}, decoration={markings,mark=at
		position .55 with {\arrow[scale=5]{>}}}] (3,-3.7) -- (4,-4.2) node[right] {$n$};
	\draw (0,-4.4) -- (2,-3.4) -- (4,-4.4);
	\draw[scalar] (3,-3.5) -- node[above left] {$\pi^0$} (4,-3);
	\draw[gauge] (4,-3) -- (5,-3.5) node[right] {$\gamma$};
	\draw[gauge] (4,-3) -- (5,-2.5) node[right] {$\gamma$};
	\end{tikzpicture}
	\caption{Diagram of the dominant background considered for the $e^+e^-$ channel, single pion production in neutral current scattering.  The neutron and other debris of the nuclear recoil must be missed and the photon pair misreconstructed as an $e^+e^-$ pair.}\label{fig:bkg-ee}
\end{figure}
\begin{figure}
	\centering
	\includegraphics[width=0.49\textwidth]{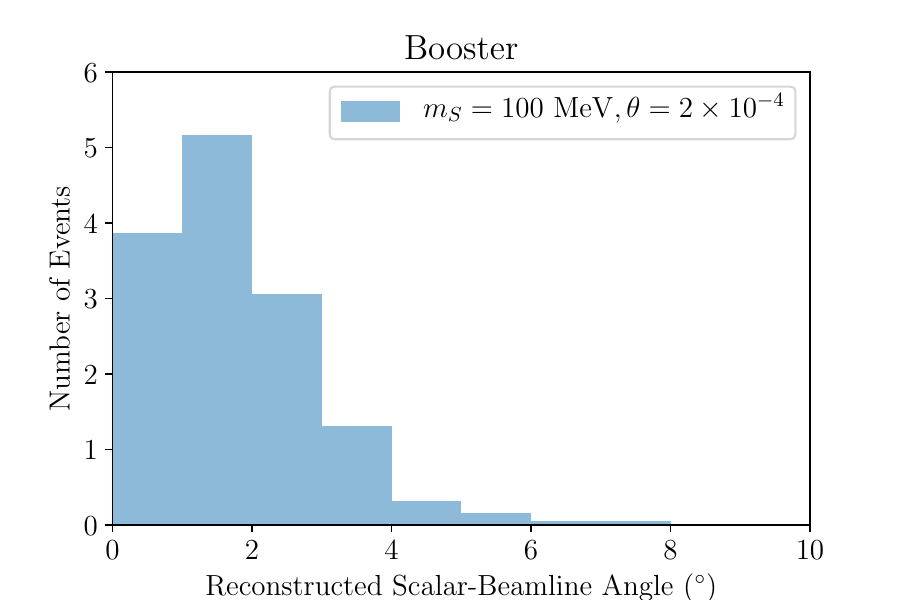}\includegraphics[width=0.49\textwidth]{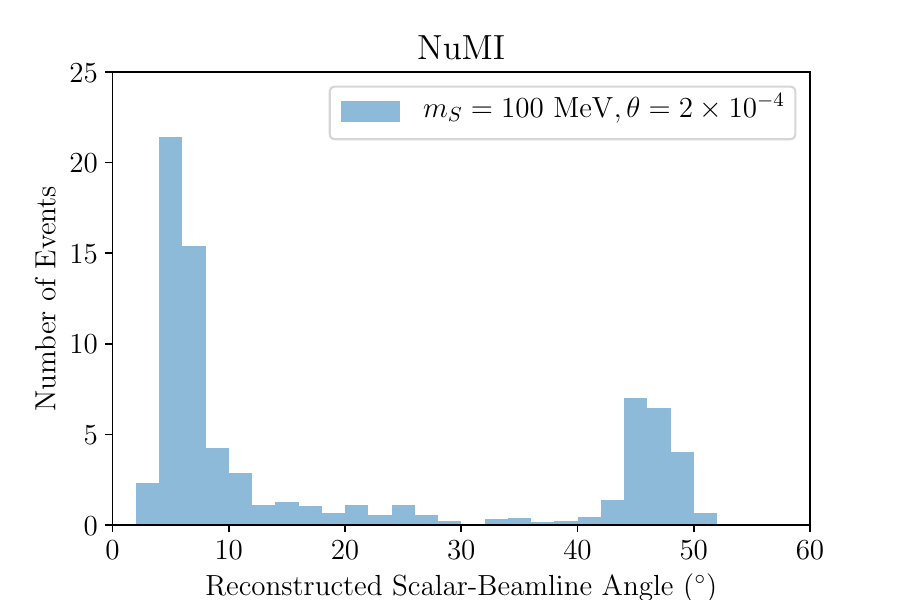}
	\caption{Distribution of the reconstructed angle of the candidate $S$ with respect to the beamline from the Booster beam at SBND and the NuMI beam at ICARUS.  The benchmark signal model here has $m_S = 100~{\rm MeV}$.  Distributions are shown using smeared candidates above threshold in events with exactly two candidate electrons, but absent any additional kinematic cuts. \label{fig:angz-ee}}
\end{figure}
\begin{figure}
	\centering
	\includegraphics[width=0.49\textwidth]{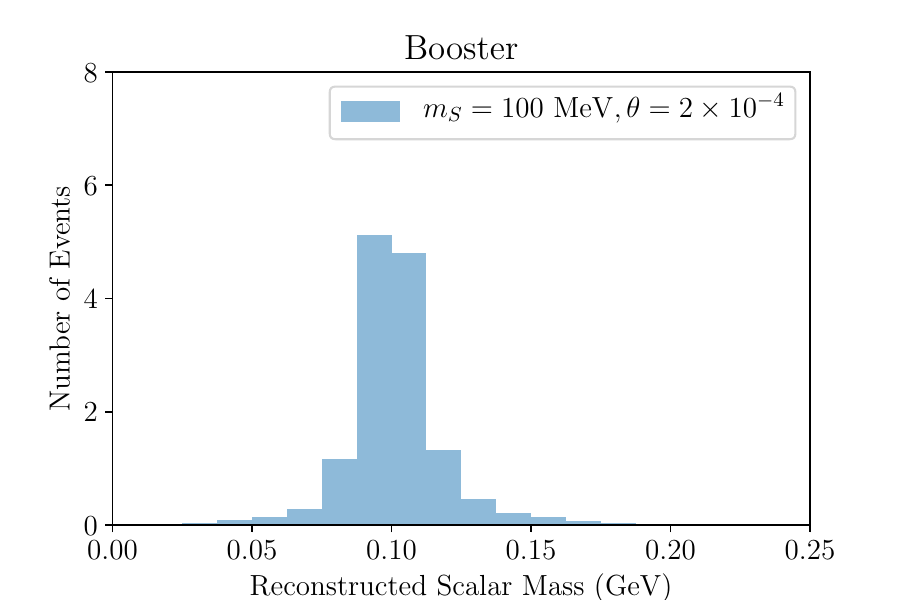}\includegraphics[width=0.49\textwidth]{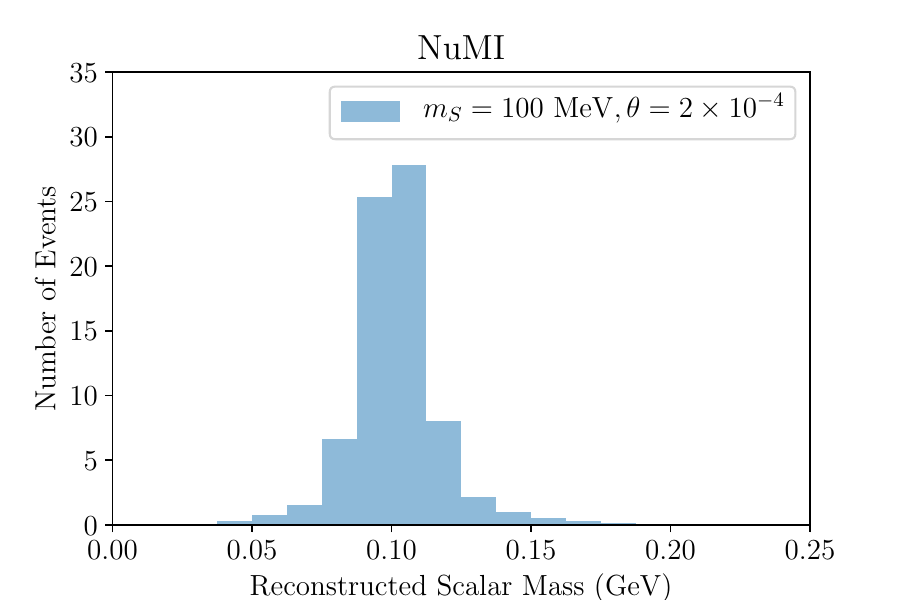}
	\caption{Distribution of the reconstructed invariant mass of the candidate $S$ from the Booster beam at SBND and the NuMI beam at ICARUS.  The benchmark signal model here has $m_S = 100~{\rm MeV}$. Distributions are shown using smeared candidates above threshold in events with exactly two candidate electrons, but absent any additional kinematic cuts.\label{fig:invm-ee}}
\end{figure}
\begin{figure}
	\centering
	\includegraphics[width=0.49\textwidth]{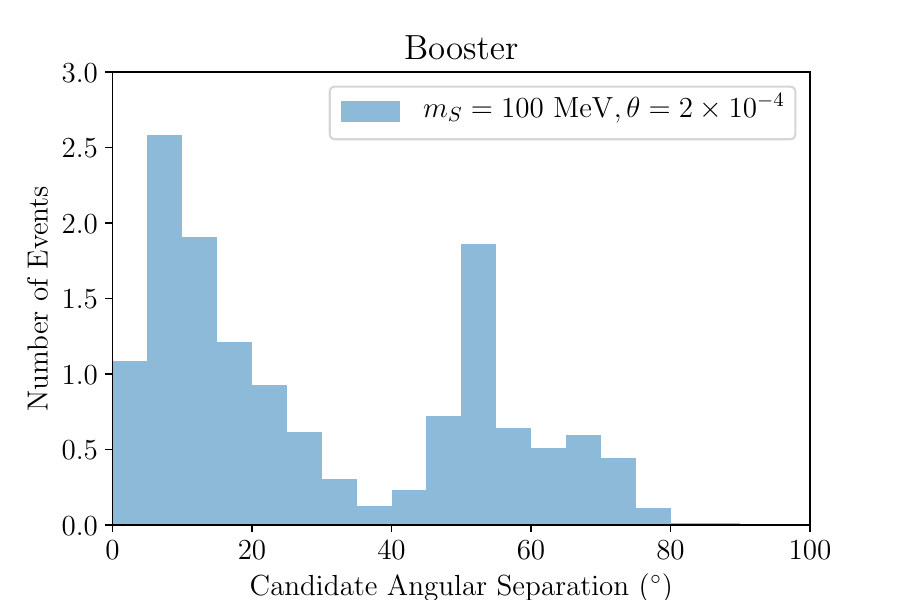}\includegraphics[width=0.49\textwidth]{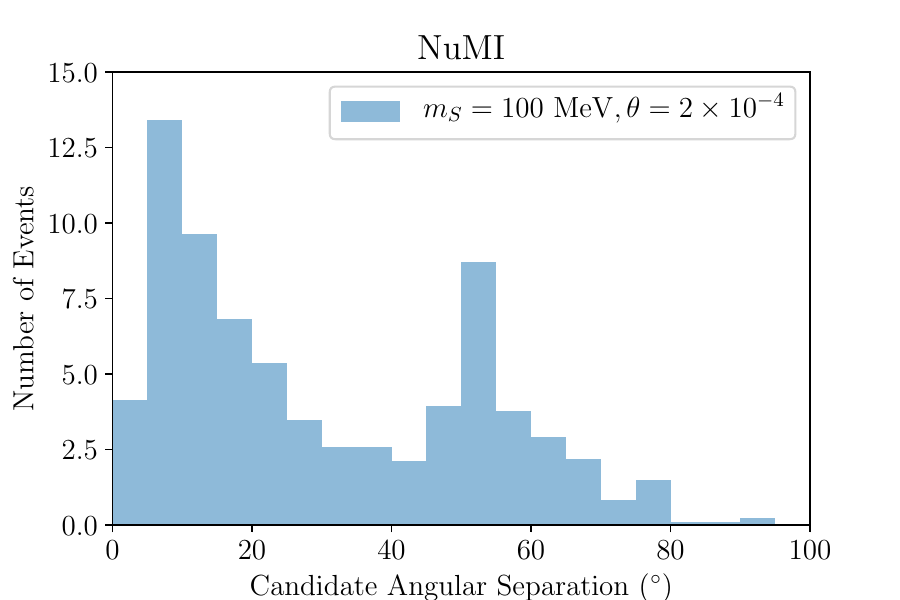}
	\caption{Distribution of the reconstructed angular separation between the candidate $S$ decay products from the Booster beam at SBND and the NuMI beam at ICARUS.  The benchmark signal model here has $m_S = 100~{\rm MeV}$. Distributions are shown using smeared candidates above threshold in events with exactly two candidate electrons, but absent any additional kinematic cuts. The second, wide angle bump is coming from scalars from KDAR, which are lower energy and lead to more widely separated decay products. \label{fig:angsep-ee}}
\end{figure}

For $m_S < 2 M_\mu$,  the dominant decay channel of the dark scalar is to an $e^+e^-$ pair.  There is essentially no irreducible background to this signature, but it is somewhat challenging to determine the reducible backgrounds.  We begin by selecting events with no reconstructed muons, charged pions, or protons above the thresholds in Table~\ref{tab:reco}.  By further selecting events with exactly two electromagnetic showers, we eliminate a large fraction of the background, as we can then require that the two showers originate at the same vertex.  To estimate the effect of this requirement, we require that the signal candidates have a separation of at least $10^\circ$ so that their vertex of production is well-reconstructed.  This is sufficient to effectively eliminate all of the background as the conversion points for diphoton events will be well separated and single photon conversions are strongly peaked at narrow opening angle, as explained above in Sec.~\ref{sec:detector-effects}. 
A cut on the angle between the reconstructed scalar momentum and the beamline could be helpful in reducing the background from cosmics, but is likely not necessary for the $e^+e^-$ case and we do not apply such a cut.  An additional search for a bump in the invariant mass of the $e^+e^-$ would lead to a measurement of the mass of the scalar.  A diagram for a candidate background event is shown in Fig.~\ref{fig:bkg-ee}.  The event distributions in the key variables on which we cut are shown in Figs.~\ref{fig:angz-ee}, \ref{fig:invm-ee}, and \ref{fig:angsep-ee}.  The distributions are drawn after the reconstruction and smearing procedure described in Sec.~\ref{sec:detector-effects} \emph{before} applying the isolation requirement and after requiring reconstruction of a single $e^+e^-$ candidate pair.  No additional kinematic cuts are applied at this level.  Note that the distributions in candidate angular separation and reconstructed scalar-beamline angle show a second peak corresponding to kaons decaying at rest to scalars, which frequently occurs in the absorber. 

\subsubsection{$\mu^+ \mu^-$ and $\pi^+ \pi^-$}

\begin{figure}[!htb]
	\begin{tikzpicture}
	\draw[fermion] (0,0) node[left] {$\nu_\mu$} -- (2,-1);
	\draw[fermion] (2,-1) -- (4,0) node[right] {$\mu^-$};
	\draw[gauge] (2,-1) -- node[right] {$W$} (2,-3);
	\draw (0,-4) -- (2,-3) -- (4,-4);
	\draw[postaction={decorate}, decoration={markings,mark=at
		position .55 with {\arrow[scale=5]{>}}}] (0,-4.2) node[left] {$n$} -- (2,-3.2);
	\draw (2,-3.2) -- (3,-3.7);
	\draw[postaction={decorate}, decoration={markings,mark=at
		position .55 with {\arrow[scale=5]{>}}}] (3,-3.7) -- (4,-4.2) node[right] {$n$};
	\draw (0,-4.4) -- (2,-3.4) -- (4,-4.4);
	\draw[scalar] (3,-3.5) -- (4,-3) node[right] {$\pi^+$} ;
	\end{tikzpicture}
	\caption{Diagram of the dominant background considered for the $\mu^+\mu^-$ channel, charged pion production in charged current scattering.  The neutron and other debris of the nuclear recoil must be missed and the charged pion must be misreconstructed as a muon.  Given that it is very challenging to distinguish a charged pion from a muon in a LArTPC, we assume that all charged pions are misreconstructed as muons, as discussed in the main text. 
	}\label{fig:bkg-mumu}
\end{figure}
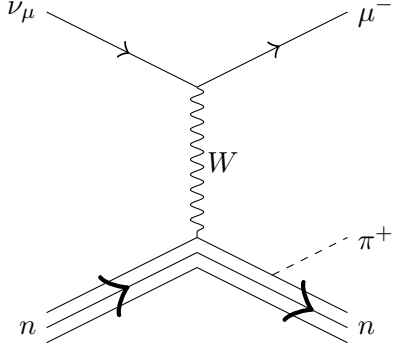
\begin{figure}
	\centering
	\includegraphics[width=0.49\textwidth]{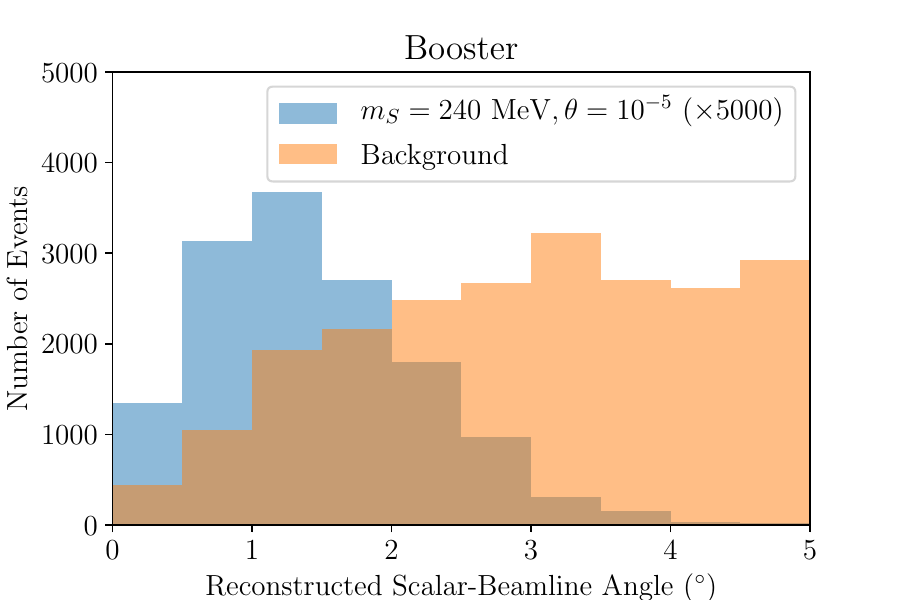}\includegraphics[width=0.49\textwidth]{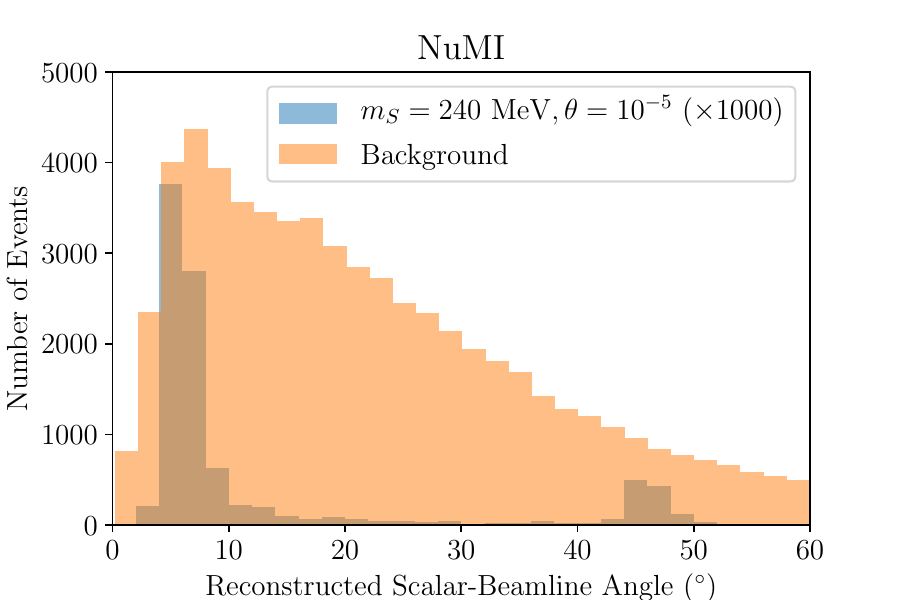}
	\caption{Distribution of the reconstructed angle of the candidate $S$ with respect to the beamline from the Booster beam at SBND and the NuMI beam at ICARUS.  The benchmark signal model here has $m_S = 240~{\rm MeV}$.  Distributions are shown using smeared candidates above threshold in events with exactly two candidate muons, but absent any additional kinematic cuts.\label{fig:angz-mumu}}
\end{figure}
\begin{figure}
	\centering
	\includegraphics[width=0.49\textwidth]{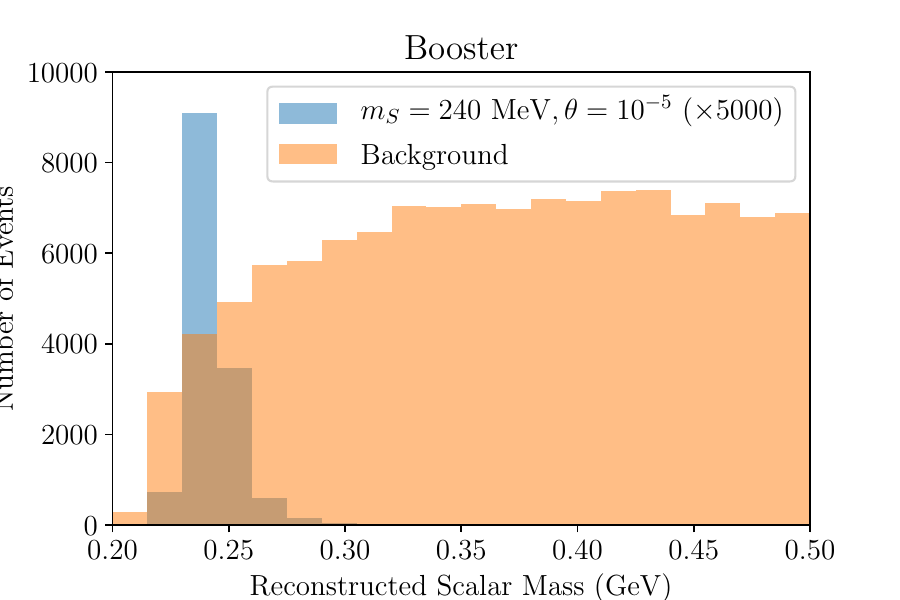}\includegraphics[width=0.49\textwidth]{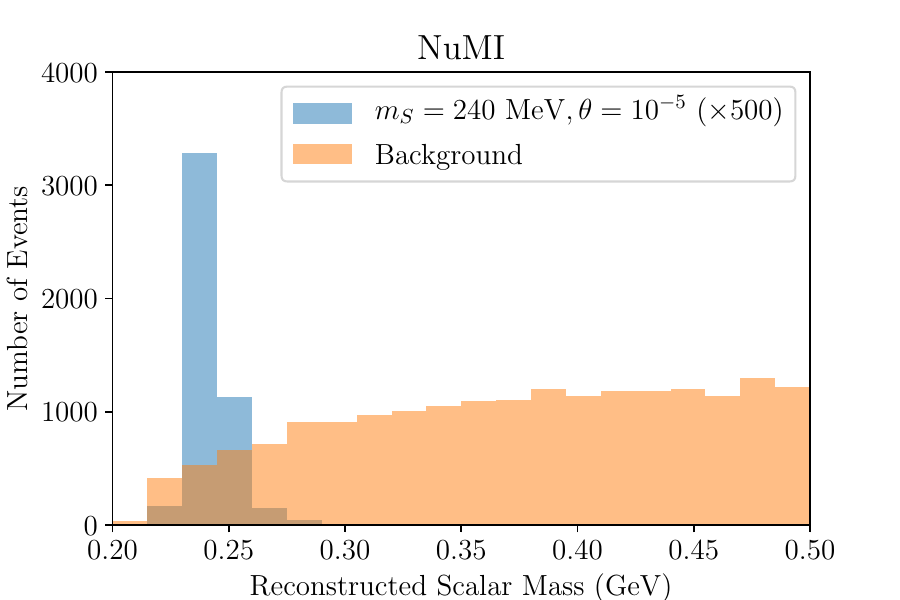}
	\caption{Distribution of the reconstructed invariant mass of the candidate $S$ from the Booster beam at SBND and the NuMI beam at ICARUS.  The benchmark signal model here has $m_S = 240~{\rm MeV}$.  Distributions are shown using smeared candidates above threshold in events with exactly two candidate muons, but absent any additional kinematic cuts.\label{fig:invm-mumu}}
\end{figure}
\begin{figure}
\centering
\includegraphics[width=0.49\textwidth]{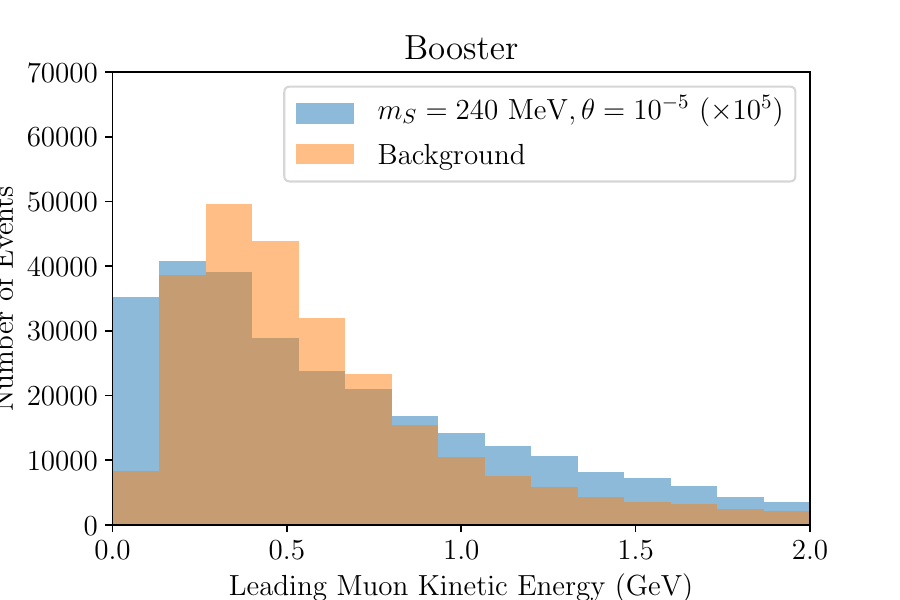}\includegraphics[width=0.49\textwidth]{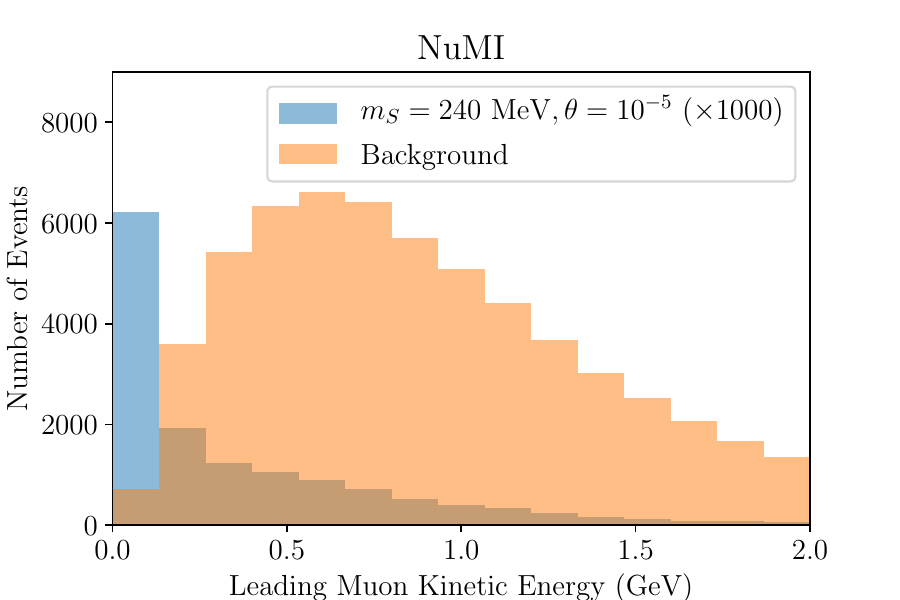}
\caption{Distribution of the leading muon candidate energy from the Booster beam at SBND and the NuMI beam at ICARUS.  The benchmark signal model here has $m_S = 240~{\rm MeV}$.  Distributions are shown using smeared candidates above threshold in events with exactly two candidate muons, but absent any additional kinematic cuts.\label{fig:maxmu-mumu}}
\end{figure}
Above $210~\text{MeV}$, muon and pion channels quickly dominate.  As discussed above, distinguishing $\mu^\pm$ and $\pi^\pm$ is difficult.  There is some recent work along this direction using machine learning techniques, for example in Ref.~\cite{Esquivel:2018fls}.  Since the ability to distinguish these particles is not fully tested, we do not attempt to do so.  We rather assume $\mu^\pm$ and $\pi^\pm$ are indistinguishable.  If they could be distinguished, it is worth noting that there is essentially no irreducible $\mu^+\mu^-$ background, so significant gains could be made.  We veto events with any reconstructed EM shower or proton above the kinetic energy thresholds in Table~\ref{tab:reco}.  We demand that there are exactly 2 candidate $\mu^\pm$/$\pi^\pm$ tracks.  A cut of $E > 2.5~{\rm GeV}$ is imposed on reconstructed scalars as we determined that this helps to separate signal from background for target-produced scalars.  We place an angular cut on the reconstructed scalar of $4^\circ$ with respect to the beamline. 
We lastly place a cut on the invariant mass of the reconstructed scalar of $40~\text{MeV}$ around the scalar mass. While the scalar mass is not known a priori, a scan over different scalar mass windows could be performed as we do in this study. A diagram for a candidate background event is shown in Fig.~\ref{fig:bkg-mumu} and the event distributions in the key variables on which we cut are shown in Figs.~\ref{fig:angz-mumu} and \ref{fig:invm-mumu}.  As in the $e^+e^-$ case, the distributions are drawn after the reconstruction and smearing procedure described in Sec.~\ref{sec:detector-effects} and after requiring reconstruction of a single $\mu^+\mu^-$ candidate pair.  No additional kinematic cuts are applied at this level. 

\subsubsection{Kaon decay at rest}
One intriguing possibility using the NuMI beamline is to make use of the fact that the detectors are off of the beam axis.  The portion of the beam that does not interact with the target is absorbed farther down the beamline.  The line between the absorber and ICARUS is about 110 m long and forms an angle of about $46^\circ$ with respect to the beamline. This raises the possibility of searching for particles coming from kaons decaying at rest (KDAR) in the absorber. In fact, MiniBooNE has recently measured monoenergetic muon neutrino charged current events arising from KDAR in the NuMI absorber~\cite{Aguilar-Arevalo:2018ylq}, and one can also envision searching for exotic particles such as dark scalars in this manner. 
 These scalars would be monochromatic and all coming from a specific angle, leading to significant background reduction.  Since the $e^+e^-$ analysis above is already effectively background free, we do not search for KDAR specifically in our analysis.  If the cosmic ray background proves to be more challenging than anticipated, searching for KDAR kinematics could prove helpful.  Furthermore, if a discovery of an excess is made, searching for KDAR would be an excellent tool for validation, as well as for probing the nature of the model due to the specific angle of $46^\circ$ for the reconstructed scalar and the monochromatic energy of the scalar.  For $\mu^+\mu^-$ and $\pi^+\pi^-$ channels, the particles coming out of the monochromatic scalar decay would be very soft and challenging to reconstruct as they are produced near threshold in the scalar decay and the scalar itself is produced close to threshold at high mass.  One could perhaps alleviate the reconstruction challenge by doing a simultaneous reconstruction of the two short candidate tracks, but development of such an analysis is beyond the scope of this work.  Due to these challenges, we do not pursue the KDAR channel further for the $\mu^+\mu^-$ or $\pi^+\pi^-$ channels either.

\subsubsection{Other channels: $\pi^0 \pi^0$, $\gamma\gamma$ }

Events with two $\pi^0$ could be rather striking, with as many as 4 EM showers.  The odds that all 4 showers are reconstructable is small and it is rather challenging to analyze such events.  It is worth noting that the branching fraction of this channel is never dominant, either.  We therefore defer further discussion of it to future analysis.

Rare decay channels of the scalar could be interesting to study as well, particularly in order to confirm the nature of the scalar in the event of a discovery.  In particular, at loop level, decay to $\gamma\gamma$ is induced just as for the standard Higgs.  Though this branching fraction is at the $10^{-3}$ level, it is of great interest for confirming the nature of the scalar.  Furthermore, non-minimal models could have a larger branching fraction to photons.  For example, in a model with heavy charged states that do not couple to the Higgs, but do couple to the dark scalar, the diphoton coupling could be larger.  

\subsubsection{Discrimination with timing}

\begin{figure}
	\centering
\includegraphics[width=0.7\textwidth]{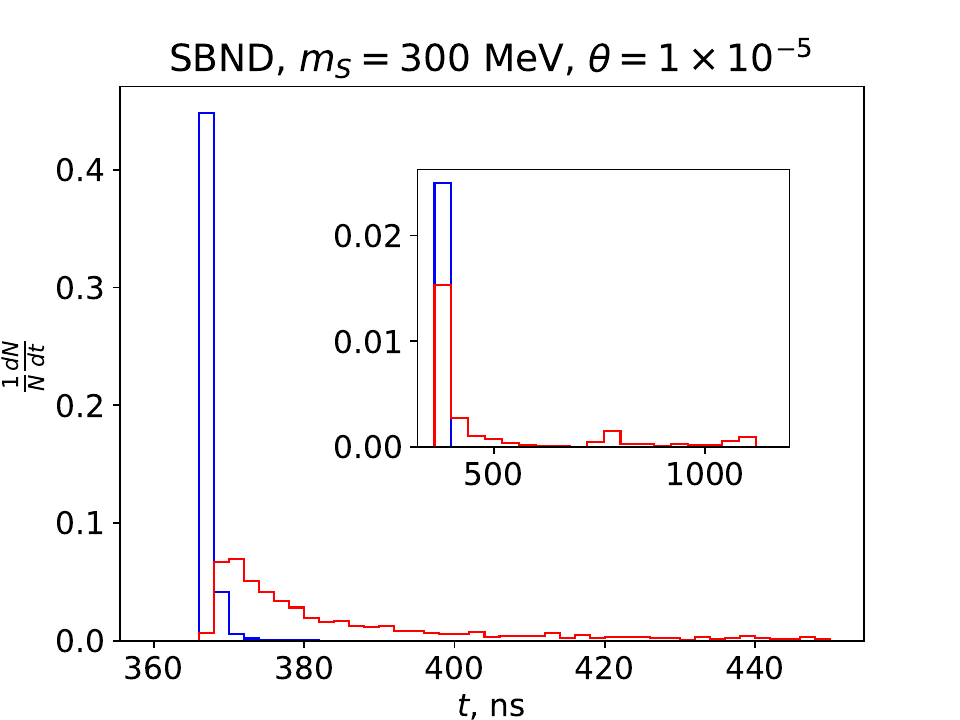}
	\caption{Timing of the scalar signal (red) and neutrino background (blue) arrival at the front of the SBND detector, relative to the initial proton collision from the Booster beam. The signal is delayed due to the massive scalar traveling between its production point and the detector. A benchmark with a 300 MeV scalar decaying to muons is shown. The inset plot shows two KDAR peaks at hundreds of ns past the initial collision, one from kaons decaying in the target and one from kaons decaying in the absorber.\label{fig:timing}}
\end{figure}

We focus in this paper on the prospects for scalar decays during the times when neutrino events would also be occurring in the SBN detectors. However, unlike neutrinos, the scalar may travel at a speed sufficiently slow compared to the speed of light that its decay is delayed relative to the neutrino arrival. While the beam spills themselves are relatively long, $1.6~\mu\text{s}$ long at BNB and $10~\mu\text{s}$ at NuMI, each spill contains multiple shorter bunches of protons. For instance, BNB employs proton bunches that are approximately 2 ns long, separated by 20 ns. The ns-level resolution of the SBN detectors can in principle be used to search for scalars decaying in between the times when neutrino events are expected~\cite{Antonello:2015lea}. We show the timing expected for the signal as compared to the neutrino background in Fig.~\ref{fig:timing} for a long-lived 300 MeV scalar. Most of the signal events occur late enough relative to the neutrino background that they could be discriminated using timing cuts. However, the delay can be much longer than the bunch spacing. There are also additional peaks at late times coming from KDAR-induced scalars that can be produced in either the target or the absorber.

It would be interesting to further investigate the use of timing to look for delayed scalar decays, as has been discussed in the context of other light hidden sector models~\cite{Ballett:2016opr,Fischer:2019fbw}. Depending on the mass, the length of the delay could vary considerably, and some care would be necessary in the case when scalars produced from one proton bunch decay in a detector after neutrino events from a later proton bunch. Furthermore, in order to perform such a study, new developments would be required in the data acquisition and triggering of the SBN experiments in order to look outside of standard timing windows.  Such a detailed study is beyond the scope of this work, and we do not attempt to use timing information further.  Nevertheless, it should prove useful in further discriminating between BSM signals from heavy particles and neutrino events.

%% file: results.tex
% !TEX root = sbnds.tex

\begin{figure}[t]
\begin{center}
\includegraphics[width=3in]{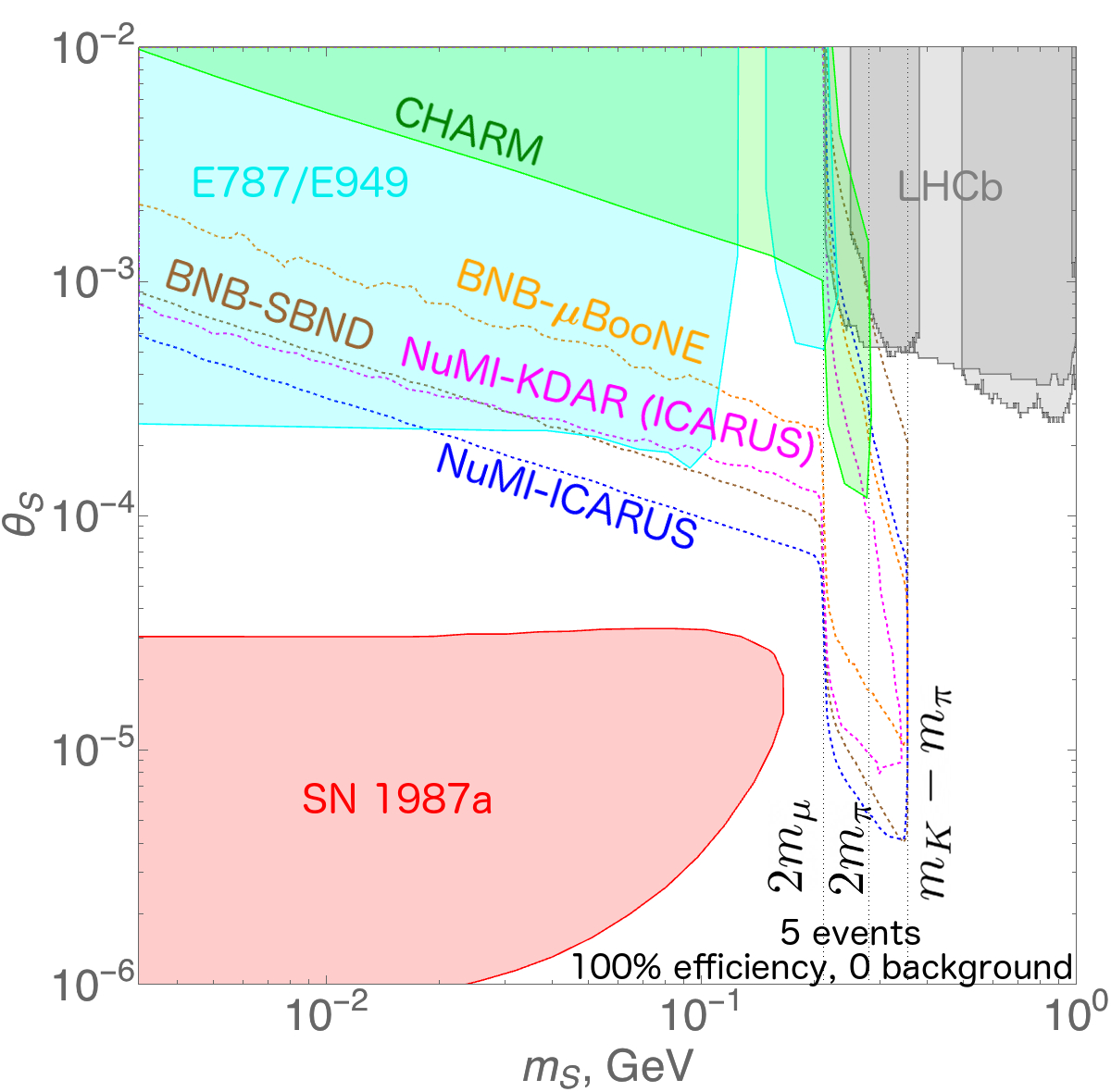}
\includegraphics[width=3in]{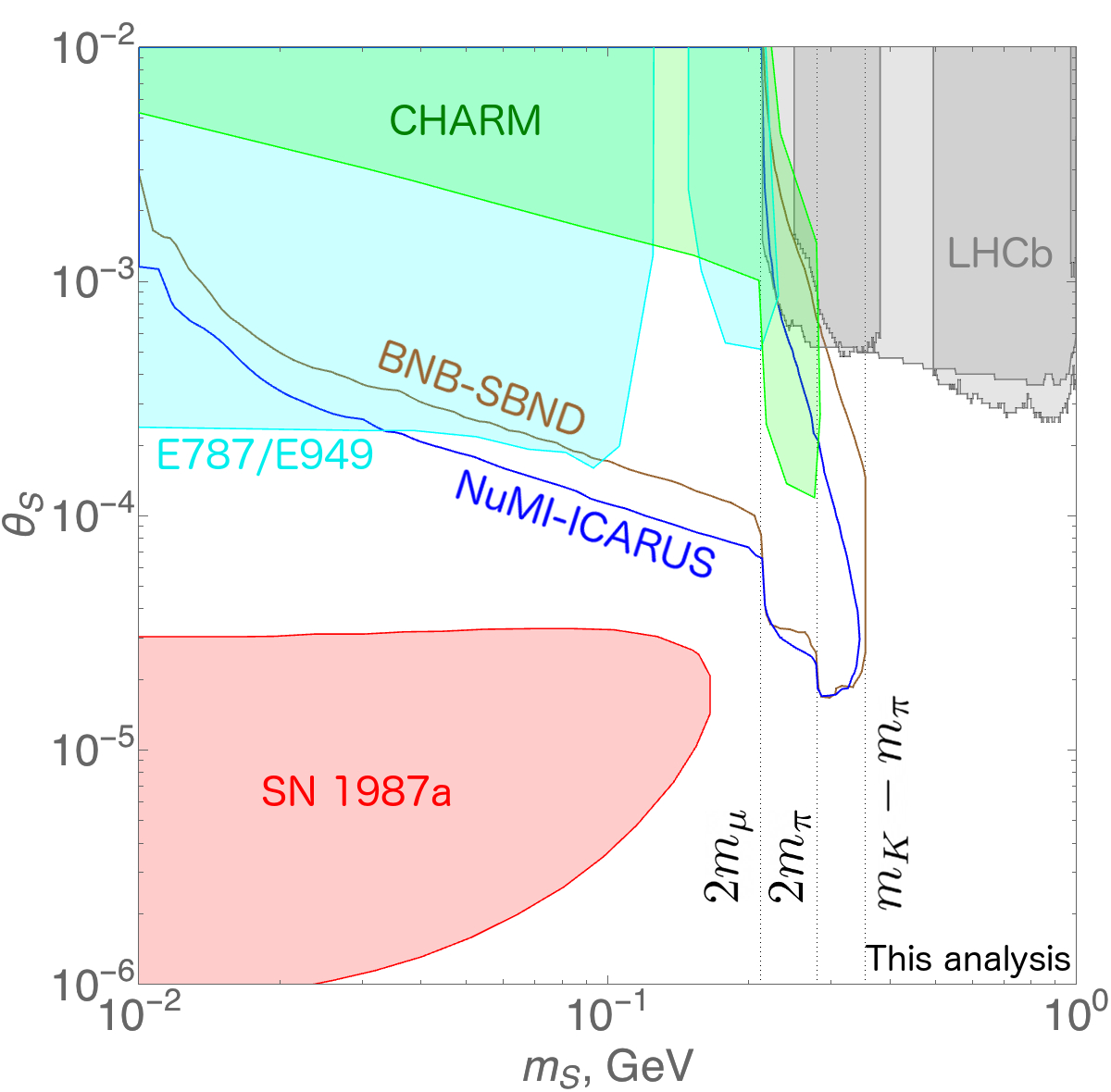}
\caption{
Projections for the on-axis SBND (brown) and MicroBooNE (orange), off-axis ICARUS (blue), and KDAR (magenta) analyses. In the left plot, dashed curves indicate 5 signal events, assuming perfect efficiencies and no backgrounds. The right plot shows the results of our analysis including estimated efficiencies and backgrounds. LHCb (gray), $K \to \pi + \mathrm{invisible}$ (cyan), and CHARM (green) limits are shown for comparison.
}
\label{figure:MB-plot}
\end{center}
\end{figure}

We display the combined results of the analyses of the previous section in Fig.~\ref{figure:MB-plot}. The regions enclosed by solid lines would be probed using our cuts, where we have conservatively assumed that sensitivity will require a minimum signal of 5 total events or 100\% of the background after all cuts are applied. We have also indicated, with dashed lines, where 5 events would occur in SBND and MicroBooNE, from Booster protons, and ICARUS, from NuMI protons. These should be thought of as the ultimate possible reach for these analyses, if efficiencies could be improved and backgrounds eliminated. Similarly, we have shown where 5 scalars would decay in ICARUS from kaons decaying at rest in the NuMI absorber, as the maximum achievable sensitivity of a targeted KDAR analysis. For comparison, we also show other current limits on light Higgs portal scalars from CHARM~\cite{Bergsma:1985qz,Winkler:2018qyg} (this bound takes into account kaon absorption in the target), LHCb~\cite{Aaij:2015tna,Aaij:2016qsm}, E787/E949~\cite{Artamonov:2009sz,Winkler:2018qyg} and SN 1987A~\cite{Krnjaic:2015mbs}. The CHARM and LHCb regions are excluded at the 95\% confidence level, while the E787/E949 region is excluded at 90\%. The SN 1987A bound is an order of magnitude estimate.

%
%%%%%%%%%% FIGURE %%%%%%%%
%%%%%%%%%%%%%%%%%%%%%%%%%%

Fig.~\ref{figure:MB-plot} shows that the SBN detectors would be sensitive to new areas of Higgs portal parameter space in the MeV-GeV region, including the gap in the E787/949 kaon decay search due to $K \to \pi \pi$ backgrounds. In addition, the Fermilab facilities would afford a better reach than the similar search at the CHARM detector, which used the 400 GeV SPS proton beam. While SPS produced many more $B$ mesons per POT than either of the Fermilab beams we consider, there were only $2.4 \times 10^{18}$ POT in all, orders of magnitude less than can be achieved at Fermilab. Also, the thick copper target used for the SPS beam led to many kaons being stopped before decaying, significantly reducing the potential number of scalars from kaon decays~\cite{Winkler:2018qyg}. Notably, off-axis production at the SBN detectors from NuMI can lead to even stronger bounds than on-axis production from the BNB. This is again primarily due to the larger number of POT that the NuMI beam can provide.

There are several proposed or upcoming facilities that could also be sensitive to light Higgs portal scalars~\cite{Beacham:2019nyx}. Those with relatively similar time scales and bounds to the SBN program include FASER~\cite{Feng:2017vli,Ariga:2019ufm}, NA62~\cite{Krnjaic:2015mbs,Izaguirre:2016dfi,NA62:2017rwk}, and SeaQuest~\cite{Aidala:2017ofy,Berlin:2018pwi}. In particular, a recent study considering scalars decaying outside NA62 suggests significant improvement is possible for scalars lighter than the dimuon threshold~\cite{Bondarenko:2019vrb}. Nevertheless, for some scalar masses we project that Fermilab facilities can test scalars with mixing angles as low as a few times $10^{-5}$, in a region which is only otherwise covered by detectors with longer times to construction and data-taking~\cite{Beacham:2019nyx}. For $m_S$ of order of hundreds of MeV, this represents an order of magnitude improvement relative to the current bounds. Owing to the large number of POT that can be collected at the Booster and NuMI targets, the SBN detectors are well-situated to be the leading probes of the Higgs portal for sub-GeV scalars.

%% file: outlook.tex
% !TEX root = sbnds.tex

Searches for sub-GeV hidden sectors have progressed rapidly in recent years, with implications for models of light dark matter, neutrino masses, and beyond. Intensity frontier experiments provide useful probes of these sectors, and in this work we have investigated the use of the SBN detectors at Fermilab to test the scalar portal. While the detectors are aligned with the 8 GeV Booster proton beam, the 120 GeV NuMI protons can also cause appreciable off-axis production, particularly at ICARUS. The sensitive LArTPC detectors are well-suited to observe the scalar decay products, including electrons, muons and pions.

Using \verb+Geant4+-based simulations of the beamlines and \verb+GENIE+ to generate neutrino-induced backgrounds, we have analyzed potential signal channels using simple kinematic cuts. As the performance characteristics of the detectors are still under investigation, we have taken generally conservative assumptions in projecting sensitivity curves, which are summarized in Figure~\ref{figure:MB-plot}. Both on-axis and off-axis production of light scalars can be seen at the SBN detectors, which achieve greater sensitivity than existing limits from CHARM, LHCb, and E787/E949. Our results show an improvement of the sensitivity to the scalar-Higgs mixing angle by over an order of magnitude for masses of order of hundreds of MeV, up to the threshold $m_K - m_\pi$. Should our estimated sensitivities be reached, the SBN detectors would provide better probes of the scalar portal than currently planned experiments on similar timescales~\cite{Beacham:2019nyx}.

With greater understanding of the background, the ``ideal'' curves in Figure~\ref{figure:MB-plot} could be attained. One could also consider running in beam dump mode to reduce neutrino background, as was done to search for dark matter in MiniBooNE~\cite{Aguilar-Arevalo:2017mqx,Aguilar-Arevalo:2018wea}. For $m_S < 2 m_e$, being able to observe nearly collinear electron-positron pairs while maintaining good separation between electrons and photons would be useful. For $m_S > 2 m_e$, it will be necessary to improve the discrimination of muons from pions. Being able to reconstruct softer objects can also help our analyses, particularly in the KDAR case where gaps appear at masses corresponding to scalars decaying to slowly moving daughter particles. Finally, we have mentioned the use of timing information to reduce the background by looking for scalars which arrive late in the detectors. It would be useful to perform a more complete analysis of this technique, including out-of-time backgrounds such as cosmic rays, in the future.

Given the small number of renormalizable portals between the SM and a new hidden sector, models for light weakly coupled mediators are highly predictive. The upcoming SBN program at Fermilab will provide a competitive probe of the Higgs portal.

%% file: validation.tex
% !TEX root = sbnds.tex

In this appendix, we validate the results of our simulation as described in Sec.~\ref{sec:simulation}, by reproducing predictions for neutrino fluxes from both the Booster and NuMI beams. The \verb+g4bnb+ and \verb+g4numi+ codes simulate the production of particles in fixed-target proton collisions, as well as their subsequent interactions and decays. When a neutrino is produced in a decay, information is stored about its production position and momentum, as well as that of the particles in its ancestry. Each neutrino is given an importance weight, which accounts for generator choices (such as a raw weighting towards higher momentum mesons produced in primary collisions for statistical purposes) as well as branching ratios.

For our analysis, we have used the neutrinos produced by \verb+g4bnb+ and \verb+g4numi+ which have kaons as parents. We generate scalars by starting with the kaon position and momentum, and simulating an isotropic 2-body decay, i.e.~$K \to \pi S$. Accounting for the relative branching fractions of $K \to \nu + \mathrm{X}$ and $K \to \pi S$, we scale the importance weight of each scalar from the neutrino weight given by the Geant 4-based simulations. Then, we check whether each randomly generated scalar would pass through the detector in question, and multiply the weight by an additional factor as in Eq.~\ref{eq:decayprob} to include the chance that the scalar decays within the detector.

\begin{figure}[t]
\begin{center}
\includegraphics[height=2.5in]{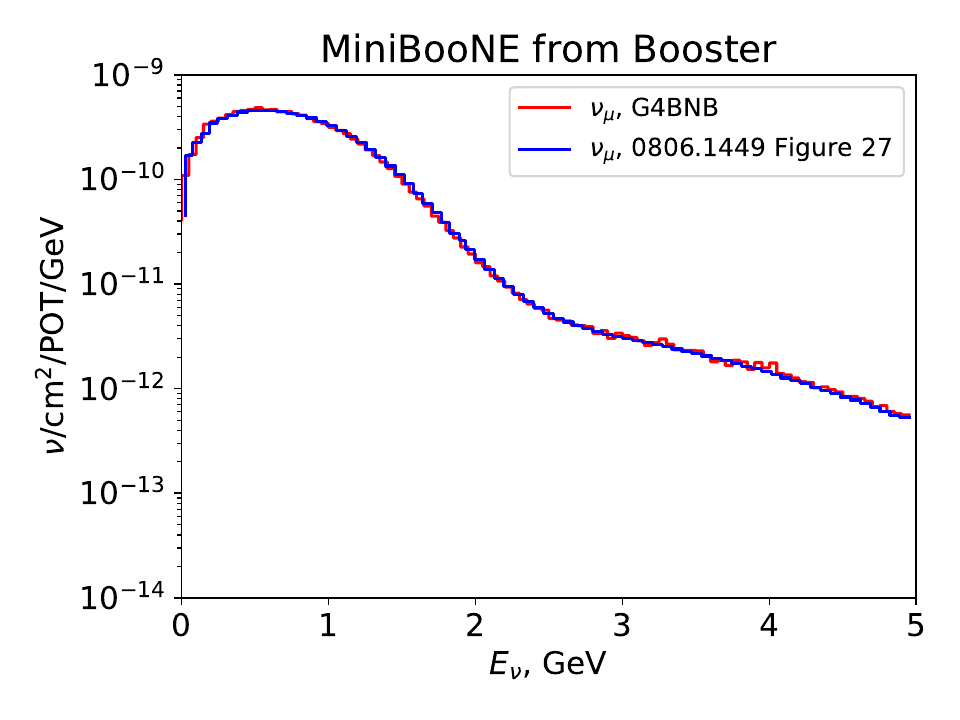}~
\includegraphics[height=2.5in]{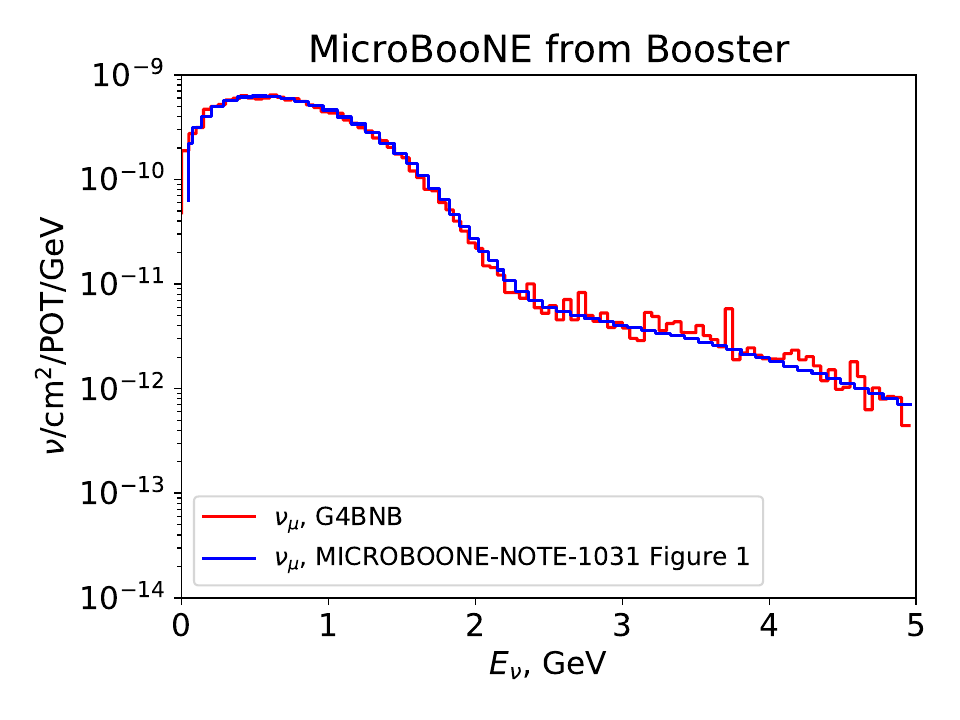}\\
\includegraphics[height=2.5in]{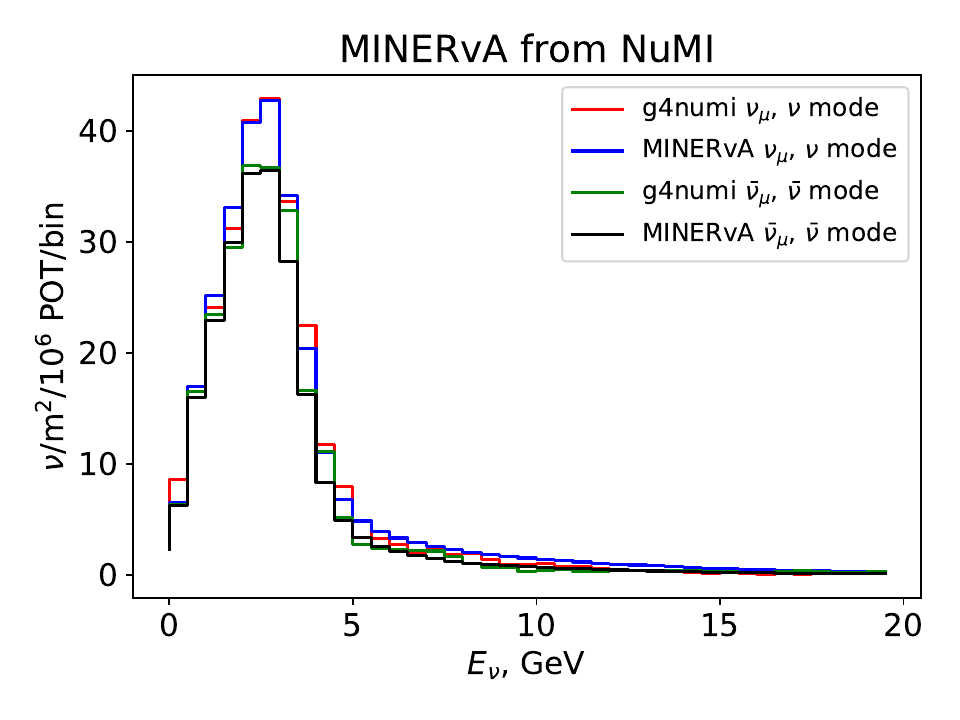}~
\includegraphics[height=2.5in]{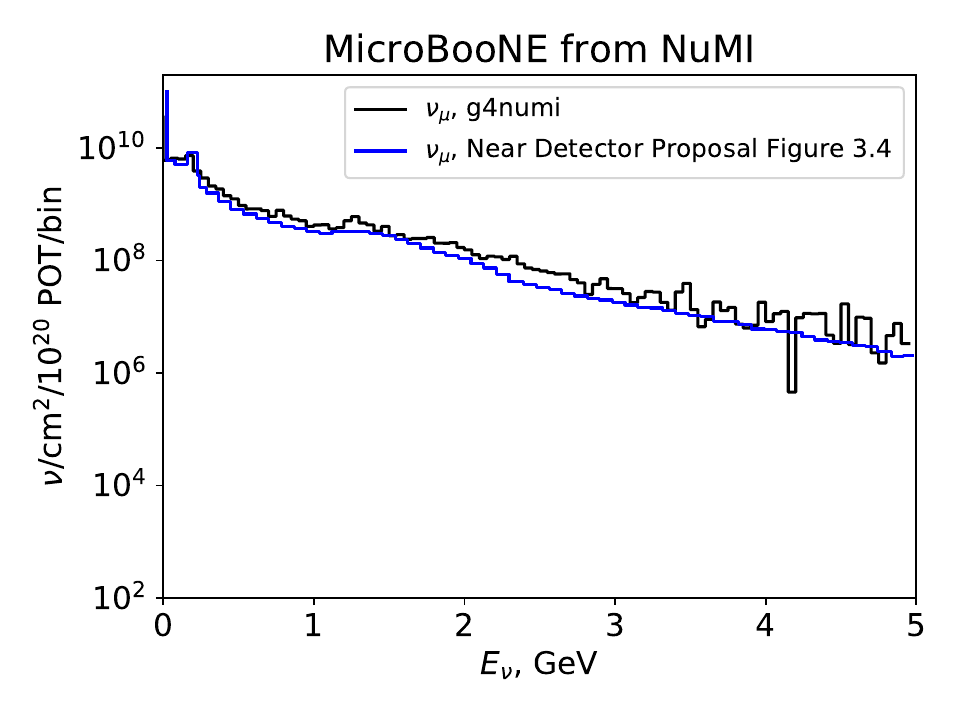}
\caption{
{Plots validating our simulation, including (top left) the flux from Booster at MiniBooNE~\cite{AguilarArevalo:2008yp}; (top right) the flux from Booster at MicroBooNE~\cite{uboonenote}; (bottom left) the flux from NuMI at MINERvA~\cite{Aliaga:2016oaz}; and (bottom right) the flux from NuMI at MicroBooNE~\cite{Admas:2013xka}.}
}
\label{fig:validation}
\end{center}
\end{figure}

To test our framework, in Fig.~\ref{fig:validation} we have used all the neutrinos coming from mesons in the Geant 4-based simulations, i.e.~pions and kaons but not muons. Then, we have replaced the $K \to \pi S$ decay with $M \to \pi \nu$, in effect reproducing the last step of the Geant 4-based neutrino simulations where a neutrino parent is decayed to a neutrino. We have then asked whether each neutrino would pass through various detectors, aiming to reproduce the Booster fluxes predicted at MiniBooNE~\cite{AguilarArevalo:2008yp} and MicroBooNE~\cite{uboonenote}, as well as the NuMI fluxes predicted on-axis at MINERvA~\cite{Aliaga:2016oaz} and off-axis at MicroBooNE~\cite{Admas:2013xka}. The 3-body muon decays which we neglect contribute negligibly to the fluxes of the neutrino flavors which we use for validation.

In all cases we see very good agreement, as most of the (anti-)muon neutrinos in (anti-)neutrino mode come from two body decays of pions or kaons, and our ignorance of muon decays and three body decays is of little consequence for the fluxes we seek to obtain. Fig.~\ref{fig:validation} confirms the validity of our procedure for extracting data about mesons from neutrinos as reported by \verb+g4bnb+ and \verb+g4numi+, as well as for simulating their decays.